\documentclass[a4paper]{jpconf-GA}

\usepackage{amsmath}
\usepackage{amssymb}
\usepackage{phonetic}

\usepackage{epsfig}
\usepackage{graphicx}

\usepackage{cite}
\numberwithin{equation}{section}

\newcommand{\be}{\begin{equation}}
\newcommand{\ee}{\end{equation}}
\newcommand{\bea}{\begin{eqnarray}}
\newcommand{\eea}{\end{eqnarray}}
\newcommand{\bra}{\langle}
\newcommand{\ket}{\rangle}
\newcommand{\sgn}{\mbox{sgn}}
\newcommand{\im}{{\mathrm{Im}}}
\newcommand{\re}{{\mathrm{Re}}}

\newcommand{\om}{\omega}

\newcommand{\pv}{{\mathbf p}}

\newcommand{\xv}{{\mathbf x}}
\newcommand{\yv}{{\mathbf y}}
\newcommand{\hm}{\hspace*{-0.5cm}}

\newcommand{\half}{\frac{1}{2}}

\newcommand{\eps}{\epsilon}
\newcommand{\vareps}{\varepsilon}

\newcommand{\bean}{\begin{eqnarray*}}
\newcommand{\eean}{\end{eqnarray*}}
\newcommand{\nn}{\nonumber}

\newcommand{\vecp}{{\mathbf p}}

\newcommand{\vecnul}{{\mathbf 0}}
\newcommand{\rmR}{{\rm R}}
\newcommand{\rmI}{{\rm I}}
\newcommand{\id}{1\!\!1}

\newcommand{\dd}{\mbox{\hausad}}

\begin{document}

\title{Introductory lectures on lattice QCD at nonzero baryon number}

\author{Gert Aarts}

\address{Department of Physics, College of Science, Swansea University, Swansea SA2 8PP, United Kingdom}

\ead{g.aarts@swan.ac.uk}

\begin{abstract}
These lecture notes contain an elementary introduction to lattice QCD at nonzero chemical potential. 
Topics discussed include chemical potential in the continuum and on the lattice; the sign, overlap and Silver Blaze problems; the phase boundary at small chemical potential; imaginary chemical potential; and complex Langevin dynamics. An incomplete overview of  other approaches is presented as well. These lectures are meant for postgraduate students and postdocs with an interest in extreme QCD. A basic knowledge of lattice QCD is assumed but not essential. Some exercises are included at the end.

\vspace*{0.1cm}
\noindent
\scriptsize{Based on lectures delivered at the {\em XIII International Workshop on Hadron Physics}, Brazil, March 2015. }

\end{abstract}

\section{Introduction}

Quantum Chromodynamics (QCD), describing the interaction between quarks and gluons, is formulated using only a few ingredients: it is an SU(3) gauge theory with six flavours of quarks, of which the lightest two are nearly massless. Yet, this results in an extremely rich theory, containing asymptotic freedom, confinement, chiral symmetry breaking, phase transitions, etc. 

\begin{figure}[h]
\centerline{ \includegraphics[height=5.8cm]{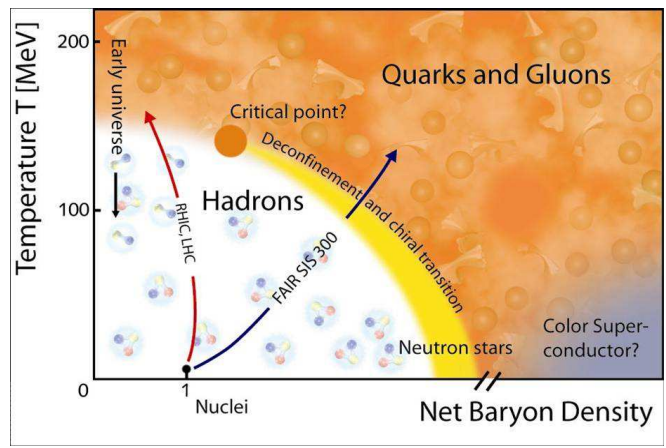}}
\caption{Sketch of the QCD phase diagram in the plane of temperature and baryon density.}
\label{fig:pd}
\end{figure}

Because QCD is strongly interacting at low energies, many interesting questions are not easily answerable. The questions relevant for these lectures concern the QCD phase diagram, i.e.\ the phase structure of strongly interacting matter, of which a sketch is presented in Fig.\ \ref{fig:pd}.
Several phases are shown: the hadronic phase at low temperature and density, the quark-gluon plasma at high temperature, possible colour-superconducting phases for cold dense matter, and perhaps there exist other phases not indicated in this version of the phase diagram. 

However, these lectures will not be about the phase diagram as such. Many good reviews are available where this is described in detail, see e.g.\ Refs.\ \cite{Kogut:2004su,Yagi:2005yb,Fukushima:2010bq}.
 Instead our focus is on a practical and concrete problem, which has been around for many years: namely why has the QCD phase diagram not yet been determined from first principles, and why is the standard nonperturbative approach, lattice QCD, not  immediately applicable?  Given these limitations, we then address what can be determined, and, importantly, what progress there is in evading this {\em status quo} altogether.

By definition these lectures are incomplete, limited by time and knowledge. Hence one may supplement them with e.g.\ the review talks presented at the annual Lattice conference  \cite{deForcrand:2010ys,Gupta:2011ma,Levkova:2012jd,Aarts:2013bla,Gattringer:2014nxa,Sexty:2014dxa,Borsanyi:2015axp}, which are usually quite accessible. My goal is to provide a basis for new postgraduate students and postdocs in the field and hence I have refrained from including very recent material which is still in development. This is especially relevant for the material discussed towards the end. 

These lectures are structured  as follows: 

\begin{itemize}
\item[]Sec.\ \ref{sec:qs}. Quantum statistical mechanics and the QCD partition function
\item[]Sec.\ \ref{sec:cont}. Chemical potential in the continuum
\item[]Sec.\ \ref{sec:lat}. Chemical potential on the lattice
\item[]Sec.\ \ref{sec:pq}. Phase-quenching: sign, overlap and Silver Blaze problems
\item[]Sec.\ \ref{sec:pb}.  Phase boundary at small chemical potential
\item[]Sec.\ \ref{sec:imag}. Imaginary chemical potential
\item[]Sec.\ \ref{sec:CL}. Complex Langevin dynamics
\item[]Sec.\ \ref{sec:CL2}. Complex Langevin dynamics for gauge theories
\item[]Sec.\ \ref{sec:other}. Other approaches
\item[]Sec.\ \ref{sec:conc}. Conclusion
\item[] Exercises
\end{itemize}

In the following section we start with a reminder of some basic concepts in quantum statistical mechanics and give the QCD partition function as a euclidean path integral. The sign problem at nonzero chemical potential is demonstrated.
Chemical potential in the continuum and on the lattice are introduced in Secs.\ \ref{sec:cont} and \ref{sec:lat}, respectively, both for fermionic and bosonic fields. The Silver Blaze problem is first mentioned in Sec.\ \ref{sec:cont} as well.
Some general remarks on the sign problem are contained  in Sec.\ \ref{sec:pq}, including the Silver Blaze problem from the viewpoint of the Dirac operator. 
The transition from the confined hadronic phase to the deconfined quark-gluon plasma  can be studied in detail using standard numerical methods for vanishing and small chemical potential. Given the ongoing heavy-ion collisions at the Relativistic Heavy-Ion Collider  (RHIC) at BNL and the Large Hadron Collider (LHC) at CERN, this is of high phenomenological relevance.
Methods applicable here are discussed in Sec.\ \ref{sec:pb}. At imaginary chemical potential the sign problem is absent and an intricate phase structure emerges, which is relevant for real chemical potential as well. This is discussed in detail in Sec.\ \ref{sec:imag}. 
At larger (real) chemical potential the sign problem prohibits the use of well-established methods. One approach which has seen some genuine progress in recent years is complex Langevin dynamics. In Secs.\ \ref{sec:CL} and \ref{sec:CL2} we focus on the basics, since this is still very much a topic in development. Finally, several other approaches are briefly discussed in Sec.\ \ref{sec:other}.  The Appendices contain some exercises, with solutions. A basic knowledge of lattice field theory is useful but not essential, except perhaps in Sec.~\ref{sec:CL2}.

\section{Quantum statistical mechanics and the QCD partition function} 
\label{sec:qs}

We consider QCD at nonzero temperature and baryon density. The standard approach in statistical mechanics uses the grand canonical ensemble, where the system is kept at a finite temperature $T$ in a spatial volume $V$ \cite{Kapusta:2006pm}. Conserved quantities, such as baryon number, are coupled to a chemical potential $\mu$, and the grand canonical partition function is given by
\be
 Z = \Tr e^{-(H-\mu N)/T} = e^{-F/T},
\ee
where $H$ is the hamiltonian (we use $\hbar=c=k_B=1$).
From the partition function, or free energy $F$, other thermodynamic quantities follow by differentiation with respect to $T$, $\mu$, etc.
For instance, the conserved number (density) is determined by 
\be
\bra N\ket=T\frac{\partial}{\partial\mu}\ln Z, \qquad\qquad \bra n\ket=\frac{1}{V}\bra N\ket,
\ee
and fluctuations in the number density by the susceptibility
\be
\bra \chi\ket = \frac{1}{V}\left[\bra N^2\ket-\bra N\ket^2\right] = \frac{\partial\bra n\ket}{\partial\mu}.
\ee
By studying the behaviour of these and other thermodynamic quantities as the external parameters are changed, the phase structure can be determined.
 
In QCD one may consider various conserved charges. For simplicity, let's take two flavours, up and down, with chemical potentials $\mu_u, \mu_d$.
To obtain quark number, we choose the quark chemical potentials equal, $\mu_u=\mu_d=\mu_q$, such that
 \be
 \bra n_q\ket = \bra n_u\ket + \bra n_d\ket.
 \ee
Note that baryon number is given by $\bra n_B\ket =  \bra n_q\ket/3$ and that the baryon chemical potential equals  $\mu_B=3\mu_q$. One way to think about chemical potential is that it corresponds to the free energy required to add one particle to the system. This makes the relation 
$\mu_B=3\mu_q$ obvious.

Another possibility is to consider a nonzero isospin density. In that case, the chemical potentials are chosen as $\mu_u=-\mu_d=\mu_{\rm iso}$, and the isospin density equals
 \be
 \bra n_{\rm iso}\ket = \bra n_u\ket - \bra n_d\ket.
\ee
Finally, we might be interested in the electrical charge density and take $\mu_u=\frac{2}{3}\mu_Q, \mu_d=-\frac{1}{3}\mu_Q$, such that the electrical charge density is given by
\be
\bra n_Q\ket=\frac{2}{3}\bra n_u\ket - \frac{1}{3}\bra n_d\ket.
\ee
In these lectures we are interested in nonzero quark (or baryon) density and $\mu$ will generically refer to quark chemical potential from now on.

Since QCD is strongly interacting in the regions of interest for the phase structure, i.e.\  where the transition from a hadron gas, or hadron plasma, to a quark-gluon plasma at high temperature or dense nuclear or quark matter at larger chemical potential and lower temperature occurs, it is necessary to use a nonperturbative  approach. Lattice QCD provides such an approach in principle and is extremely successful at $T>0$ and $\mu\sim 0$ (as well as of course at $T=\mu=0$).
However, a full determination of the QCD phase diagram requires numerical simulations in the entire $T-\mu$ plane and this is where the stumbling block appears. Due to what is known as the sign problem, there is at present no first-principle determination of the QCD phase diagram, the main motivation for these lectures.

On the lattice the QCD partition function is written as an euclidean path integral,\footnote{Note that I will not review the lattice formulation, see e.g.\ the textbooks \cite{Smit:2002ug,Gattringer:2010zz}.
 It suffices to know that it is formulated in terms of the links  $U_{x\nu} = e^{iaA_{x\nu}}$, with $A_{x\nu}$ the vector potential and $a$ the lattice spacing. The inverse temperature is given by the extent in the temporal direction, $1/T=a N_\tau$, with $N_\tau$ the number of time slices.
We will often use `lattice units', $a\equiv 1$.} 
\be
\label{eq:Z}
Z=\int DU D\bar\psi D\psi \, e^{-S} = \int DU\, e^{-S_{\rm YM}}\det M(\mu),
\ee
where $U$ denote the gauge links and $\psi, \bar\psi$ the quark fields. The QCD action has the following schematic form
\be
S = S_{\rm YM} + \int d^4x\, \bar\psi M\psi.
\ee
Here $S_{\rm YM}$ is the Yang-Mills action, depending on $U$, and $M$ denotes the fermion matrix, depending on $U$ and the chemical potentials. 
Integrating over the quark fields yields the right-hand side of Eq.\ (\ref{eq:Z}), which contains the determinant $\det M$. 

Now, in numerical simulations  the integrand,
\be
\rho(U) \sim e^{-S_{\rm YM}}\det M(\mu),
\ee 
is interpreted as a (real and positive) probability weight such that configurations of gauge links can be generated, relying on importance sampling.
However, at nonzero chemical potential the fermion determinant turns out to be not real and positive but complex,
\be
\label{eq:cD}
\left[\det M(\mu)\right]^* = \det M(-\mu^*) \in \mathbb C,
\ee
as is reviewed below.
As a result the weight $\rho(U)$ is complex as well and standard numerical algorithms based on importance sampling are not applicable. 
As stated above, this is usually referred to as the sign problem, even though complex-phase problem would be more accurate.
The goal of these lectures is to review these statements, as well as to discuss some partial or possibly  complete solutions to  the sign problem.

\section{Chemical potential in the continuum}
\label{sec:cont}

\subsection{Fermions}

Let us consider noninteracting fermions, with the euclidean action,\footnote{We use the following conventions:
\[
\gamma_\nu^\dagger=\gamma_\nu\quad\gamma_5^\dagger=\gamma_5,
\quad \{\gamma_\mu,\gamma_\nu\}=2\delta_{\mu\nu}, 
\quad \{\gamma_\nu,\gamma_5\}=0,
\quad\gamma_5^2=1.
\]
}

\be
S = \int_0^{1/T} \!\!\! d\tau \int d^3x \, \bar\psi\left(\gamma_\nu\partial_\nu+m\right)\psi.
\ee
Due to the global symmetry
\be
\psi\to e^{i\alpha}\psi, \qquad\quad\quad \bar\psi\to\bar\psi e^{-i\alpha},
\ee
fermion number is a conserved charge,
\be
N = \int d^3 x\,\bar\psi\gamma_4\psi=\int d^3x \, \psi^\dagger\psi \qquad\Rightarrow \qquad \partial_\tau N=0.
\ee
To obtain the grand canonical partition function in the euclidean path integral formulation, we add the following term to the action,
\be
\label{eq:muN}
\frac{\mu N}{T} = \frac{\mu}{T} \int d^3 x\,\bar\psi\gamma_4\psi=\int_0^{1/T} d\tau\int d^3x\, \mu \bar\psi\gamma_4\psi,
\ee
which, after the inclusion of an abelian gauge field $A_\nu$, now reads
\be
S =  \int_0^{1/T} d\tau\int d^3x\, \bar\psi\left[ \gamma_\nu(\partial_\nu+i A_\nu)+\mu\gamma_4+m\right] \psi
= \int d^4x\,\bar\psi M\psi.
\ee
We can now make a few observations:
 \begin{itemize}
 \item $\mu$ appears in the same way as $i A_4$, i.e.\ as the imaginary part of the four-component of an abelian vector field. This will be important when chemical potential is introduced in the lattice formulation.
 \item
The action is complex. This can be seen by the absence of  `$\gamma_5$ hermiticity'. 
At $\mu=0$ it is easy to check that
\be
\left(\gamma_5 M\right)^\dagger = \gamma_5 M, \qquad\qquad M^\dagger = \gamma_5 M\gamma_5,
\ee
leading to
\be
\det M^\dagger = \det \left(\gamma_5 M\gamma_5\right) = \det M = (\det M)^*,
\ee
i.e.\ the determinant is real. On the other hand, when $\mu\neq 0$ we find
\be
M^\dagger(\mu) = \gamma_5 M(-\mu^*)\gamma_5,
\ee
resulting in Eq.\ (\ref{eq:cD}) and a complex determinant.
\item
When the chemical potential is chosen to be purely imaginary, the determinant is real again. This has been exploited extensively and will be discussed below.
\item For abelian gauge theories, the chemical potential can be removed by a simple gauge transformation of $A_4$ (choose $\mu$ imaginary and use analyticity). This is no longer true in SU($N$) theories or for theories with more than one chemical potential.
\end{itemize}

It is important to realise that the sign problem not specific for fermions. In particular, it is not due to the Grassmann nature of fermions, since, after all, standard lattice simulations work well at zero chemical potential. Instead it arises from the complexity of the determinant (or the action) in the path integral weight, and as such it is also present in bosonic theories.

\subsection{Bosons}

To illustrate this, let us now consider a complex scalar field, with again a global symmetry $\phi\to e^{i\alpha}\phi$. The action is
\be
 S= \int d^4x \left(|\partial_\nu\phi|^2+m^2|\phi|^2+\lambda|\phi|^4\right),
\ee
and the conserved charge reads
\be
 N=\int d^3x\, i\left[\phi^*\partial_4\phi-(\partial_4\phi^*)\phi\right].
 \ee
 In order to write down the euclidean path integral at nonzero $\mu$ for this case, we have to revisit the derivation of the path integral with a bit more care than above \cite{Kapusta:2006pm}. We start from the partition function,
\be
Z = \Tr e^{-(H-\mu N)/T},
\ee
and express the hamiltonian and conserved charge (densities) in terms of the canonical momenta $\pi_1=\partial_4\phi_1, \pi_2=\partial_4\phi_2$, where $\phi=(\phi_1+i\phi_2)/\sqrt{2}$. For example, the charge now takes the form
\be
N = \int d^3x\left(\phi_2\pi_1-\phi_1\pi_2\right),
\ee
and the partition function reads 
\bea
Z =&&\hm   \Tr e^{-(H-\mu N)/T}  \nn\\
=&&\hm  \int D\phi_1D\phi_2\int D\pi_1D\pi_2 \exp\int d^4x\Big[ i\pi_1\partial_4\phi_1+i\pi_2\partial_4\phi_2
-{\cal H}+\mu(\phi_2\pi_1-\phi_1\pi_2)\Big].
\eea
After integrating out the momenta, we find the following expression for the euclidean action
\bea 
S =&&\hm \int d^4x\left[ (\partial_4+\mu)\phi^*(\partial_4-\mu)\phi+|\partial_i\phi|^2+m^2|\phi|^2+\lambda |\phi|^4\right]
\nn
\\
=&&\hm \int d^4x\left[ |\partial_\nu\phi|^2+(m^2-\mu^2)|\phi|^2+\mu(\phi^*\partial_4\phi-\partial_4\phi^*\phi)+\lambda |\phi|^4\right].
\eea
We observe that the chemical potential appears again as an imaginary vector potential.
In the second line, the linear term in $\mu$ is purely imaginary, resulting in a complex action $S^*(\mu) = S(-\mu^*)$, while the quadratic term in $\mu$ arose from integrating out the momenta and is absent in fermionic theories.

The bosonic theory is discussed in much more detail in \ref{App:A} in the form of an exercise.

\subsection{Towards the Silver Blaze problem}
\label{sec:3.3}

Consider a particle with mass $m$ and a conserved charge at low temperature: as mentioned earlier, $\mu$ is the change in free energy when a particle carrying the conserved charge is added, i.e.\ the energy cost for adding one particle.
Hence
\begin{itemize}
\item if $\mu<m$: not enough energy available to create a particle $\Rightarrow$ no change in the groundstate;
\item if $\mu>m$: plenty of energy available $\Rightarrow$  the groundstate has a nonzero density of particles.
\end{itemize}
Hence it follows from simple statistical mechanics that at zero temperature the density becomes nonzero (the `onset')  at $\mu=\mu_c\equiv m$. We will now demonstrate this for free fermions.

The standard expression for the logarithm of the partition function for a free relativistic fermion gas is given by \cite{Kapusta:2006pm}
\be
\ln Z = 2V\int \frac{d^3p}{(2\pi)^3}\left[\beta\om_\pv
+\ln\left(1+e^{-\beta(\om_\pv-\mu)}\right)
+\ln\left(1+e^{-\beta(\om_\pv+\mu)}\right)
\right],
\ee
where $\om_\pv=\sqrt{\pv^2+m^2}$ and $\beta=1/T$. The 2 arises from spin, the first term is the zero-point energy and the other terms represent particles and anti-particles at nonzero temperature and chemical potential.
The density is given by 
\be
\bra n\ket = \frac{T}{V}\frac{\partial \ln Z}{\partial\mu}
= 2\int \frac{d^3p}{(2\pi)^3} \left[ \frac{1}{e^{\beta(\om_\pv-\mu)}+1} - \frac{1}{e^{\beta(\om_\pv+\mu)}+1}\right].
\ee

Let us consider the low-temperature limit, $T\to 0$. We have to distinguish two cases:
\begin{itemize}
\item  $\mu<m$: the `1' in the denominator of the Fermi-Dirac distribution can be ignored and 
\be
   \bra n\ket\sim 2 \int \frac{d^3p}{(2\pi)^3} \left[ e^{-\beta(\om_\pv-\mu)} - e^{-\beta(\om_\pv+\mu)}\right] \to 0.
 \ee
Particles and antiparticles are thermally excited but Boltzmann suppressed.
\item
  $\mu>m$: in this case $\mu$ can be larger than $\om_\pv$ and the first Fermi-Dirac distribution becomes a step function at $T=0$,  
\be
 \bra n\ket\sim 2 \int \frac{d^3p}{(2\pi)^3}\,  \Theta(\mu-\om_\pv) = \frac{\left(\mu^2-m^2\right)^{3/2}}{3\pi^2}\Theta(\mu-m).
\ee
We find a nonzero density,  with the onset at $\mu=m$, as expected. 
\end{itemize}

At strictly zero temperature, we note therefore that thermodynamic quantities  (free energy, pressure, $\bra n\ket, \chi,\ldots$) are independent of $\mu$ when $\mu<\mu_c$, i.e.\ as long as $\mu$ is below the mass of the lightest particle in the channel with the appropriate quantum numbers. How this independence emerges in numerical simulations is nontrivial and has been dubbed the {\em Silver Blaze} problem  \cite{Cohen:2003kd},  to be discussed further below. 
Finally, we note that the same holds for bosons, see \ref{App:A} for details.

\section{Chemical potential on the lattice}
\label{sec:lat}

We now discuss how to add chemical potential to the action on the lattice. 
Naively adding $\mu\bar\psi\gamma_4\psi$, see Eq.\ (\ref{eq:muN}),  leads to $\mu$-dependent ultraviolet divergences \cite{Hasenfratz:1983ba}. However, this is not expected on general grounds, since the presence of temperature or chemical potential should not affect renormalisation at short distances.
Instead, we better follow the observations made in the continuum:
\begin{itemize}
\item the chemical potential couples to the conserved charge;
\item it appears as the imaginary part of the four-component of an abelian vector field. 
\end{itemize}
We consider a lattice action, where the derivatives are replaced by simple nearest-neighbour terms. 
The terms in the action from which the conserved lattice current follows, the so-called hopping terms,  are\footnote{Note that under a unitary gauge transformation, $\psi_x\to \Omega_x\psi, 
 \bar \psi_x\to \bar\psi \Omega_x^\dagger, U_{x\nu}\to  \Omega_x  U_{x\nu} \Omega_{x+\nu}^\dagger$ with $\Omega_x^\dagger\Omega_x=\id$; hence these terms are gauge invariant.}
\be
\label{eq:action}
 S\sim \bar\psi_xU_{x\nu}\gamma_\nu\psi_{x+\nu} - \bar\psi_{x+\nu}U^\dagger_{x\nu} \gamma_\nu\psi_x,
 \ee
 where $\nu=1,2,3,4$.
 The exactly conserved (point-split) current reads then
 \be
  j_\nu\sim  \bar\psi_xU_{x\nu}\gamma_\nu\psi_{x+\nu} + \bar\psi_{x+\nu}U^\dagger_{x\nu} \gamma_\nu\psi_x.
  \ee
  Chemical potential is now introduced  \cite{Hasenfratz:1983ba,Kogut:1983ia} as an imaginary abelian vector field in the 4-direction, i.e.\ in the temporal hopping terms,
\bea
\nn
\mbox{forward hopping:}   \qquad\quad  U_{x4} = e^{iA_{4x}} & \quad\Rightarrow\quad & e^{a\mu}, \\
\mbox{backward hopping:}  \quad\quad  U_{x4}^\dagger = e^{-iA_{4x}} & \quad\Rightarrow\quad &  e^{-a\mu},
\label{eq:hop}
\eea
where $a$ is the lattice spacing in the temporal direction, written explicitly here. It is easy to check that an expansion in small $a\mu$ yields the correct (naive) continuum limit and that the chemical potential couples to the exactly conserved charge, even for finite lattice spacing. Moreover, no new ultraviolet divergences appear. Note that different prescriptions are possible, provided that they agree in the continuum limit \cite{Bilic:1983fc}.

Note that typically expressions are written in terms of lattice units, $a\equiv 1$, and hence the factor $a$ is not included in the exponentials. However, it is always clear what is meant, since $\mu$ will appear in the following combination,  
\be
\mbox{(lattice notation)} \quad\qquad  \mu N_\tau = \mu/T \quad\qquad \mbox{(continuum notation)}.
\ee
The reason for this is explained shortly.

\begin{figure}[t]
\centerline{ 
\includegraphics[height=4cm]{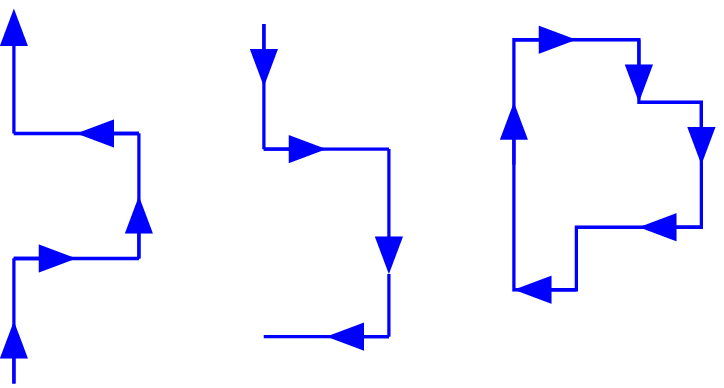}
$\quad e^{\mu N_\tau} = e^{\mu/T} $
\includegraphics[height=3cm]{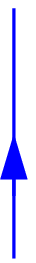}
$\quad e^{-\mu N_\tau} = e^{-\mu/T}$
\includegraphics[height=3cm]{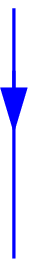} 
}
\vspace*{0.4cm}
\centerline{\hspace*{1cm}(a) \hspace*{8cm} (b)}
\caption{(a) Forward (backward) hopping is (dis)favoured by $e^{\mu n_\tau}$ ($e^{-\mu n_\tau}$), while closed loops are $\mu$-independent. (b) Loops wrapping around the temporal direction contribute $e^{\pm\mu/T}$.
}
\label{fig:hop}
\end{figure}

As can be seen in Eq.\ (\ref{eq:hop}), the chemical potential introduces an imbalance between forward and backward hopping in the euclidean-time direction, 
\bea
\nn
\mbox{$n_\tau$ steps of forward hopping (quark)} & \quad\Rightarrow\quad & \mbox{favoured as} \;\; e^{\mu n_\tau}; \\
\mbox{$n_\tau$ steps of backward hopping (anti-quark)} & \quad\Rightarrow\quad & \mbox{disfavoured as} \;\; e^{-\mu n_\tau}.
\eea
It follows that in closed worldlines the $\mu$ dependence cancels exactly, see Fig.\ \ref{fig:hop}.

Therefore the $\mu$ dependence will only survive when worldlines wrap around the time direction. This suggests that $\mu$ can effectively be thought of as a boundary condition. This notion can be made explicit as follows: consider the field redefinition 
\be
\label{eq:redef}
 \psi_x=e^{-\mu\tau}\psi_x', \qquad\qquad\quad  \bar\psi_x=e^{\mu\tau}\bar\psi_x'.
 \ee
The $\mu$  dependence then drops from all terms of the form $\bar\psi_x e^\mu\psi_{x+4}$ and  $\bar\psi_{x+4} e^{-\mu}\psi_{x}$ 
(and also from terms on the same time slice, such as spatial hopping terms), but appears instead as a boundary condition,\footnote{The minus sign is due to the anticommuting nature of fermions.} 
\be
\psi_{N_\tau}=-\psi_0 \qquad\quad\Rightarrow \qquad\quad  \psi'_{N_\tau}=-e^{\mu N_\tau} \psi'_0,
\ee
wrapping around the temporal direction. This explains the presence of the factor $e^{\pm\mu/T}$.

The Bose gas on the lattice is discussed as an exercise in \ref{App:A}.

\section{Phase-quenching: sign, overlap and Silver Blaze problems}
\label{sec:pq}

\subsection{Overlap problem}

Let us consider again the partition function
\be
Z = \int DUD\bar\psi D\psi\, e^{-S}  = \int DU \, e^{-S_B} \det M,
\ee
with a complex determinant,
\be
\det M = |\det M| e^{i\varphi}.
\ee
 An apparently straightforward solution to the complex-phase problem is to absorb the phase in the observable, as follows
\be
\bra O\ket_{\rm full} = \frac{\int DU\, e^{-S_B} \det M \, O}{\int 
DU\, e^{-S_B} \det M}
= \frac{\int DU\, e^{-S_B} |\det M| \,e^{i\varphi} O}{\int
DU\, e^{-S_B} |\det M| \,e^{i\varphi}}
= \frac{\bra e^{i\varphi} O\ket_{\rm pq}}{\bra e^{i\varphi} 
\ket_{\rm pq}}, 
\label{eq:pq}
\ee
where 
 $\bra\cdot \ket_{\rm full}$ denotes expectation values taken with respect to the original, complex weight, while
 $\bra\cdot \ket_{\rm pq}$ denotes expectation values with respect to the phase-quenched weight, i.e.\ using $|\det M|$.
Every step in Eq.\ (\ref{eq:pq}) is well defined in principle.

To analyse why this method is nevertheless not applicable in general, let us take a closer look at the average phase factor $\bra e^{i\varphi}\ket_{\rm pq}$.
It is simple algebra to write
\be
\bra e^{i\varphi} \ket_{\rm pq} = 
\frac{\int DU\, e^{-S_B} |\det M| \, e^{i\varphi} }{\int
DU\, e^{-S_B} |\det M|} = \frac{Z_{\rm full}}{Z_{\rm pq}} = 
e^{-\Omega\Delta f},
\ee
where we have expressed the partition functions  in terms of the free energy densities, 
\be
Z \equiv Z_{\rm full} = e^{-F/T} = e^{-\Omega f},
\qquad\qquad
Z_{\rm pq} = e^{-F_{\rm pq}/T} = e^{-\Omega f_{\rm pq}},
\ee
 with $\Omega$ the spacetime volume ($\Omega=V/T$ in physical units or $N_\tau N_s^3$ in lattice units),
 and 
 \be
 \Delta f= f-f_{\rm pq}
 \ee
 is the difference in the free energy densities.
 Note that $Z_{\rm full}\leq Z_{\rm pq}$.
We find therefore that the average phase factor is the ratio of two partition functions and that it goes to zero in the thermodynamic limit, unless $f=f_{\rm pq}$.
As a consequence the ratio in Eq.\ (\ref{eq:pq}) is not defined! Both numerator and denominator vanish exponentially as the spacetime volume is increased.

The reason for this is the so-called {\em overlap problem}: the phase-quenched theory is manifestly different from the full theory and hence, even though the weight in the phase-quenched theory is real and positive, sampling from it is a highly ineffective approach to mimic sampling from the full theory.
Because of the exponential dependence on the four-volume, it is often said that the sign problem is exponentially hard.

The overlap problem emphasises that the physics of the phase-quenched and the full theory differ in an essential way. This can be nicely illustrated in QCD with two degenerate flavours \cite{Son:2000xc}.
Recall that $M^\dagger(\mu)=\gamma_5 M(-\mu)\gamma_5$ for real quark chemical potential $\mu$. Then the determinant in the full theory reads
\be
\left[\det M(\mu)\right]^2
\ee
while the determinant in the phase-quenched theory can be written as
\be
 \left|\det M(\mu)\right|^2 =   \det M^\dagger(\mu) \det M(\mu) =   \det M(-\mu) \det M(\mu).
\ee
Hence the phase-quenched theory corresponds in fact to a theory at nonzero isospin chemical potential (see Sec.\ 2). It turns out that this theory has a very different phase structure than QCD at nonzero baryon chemical potential, especially at low temperature. 
This can be understood from the discussion in Sec.\ \ref{sec:3.3}: while the lightest particle with nonzero baryon number is the nucleon, the lightest particle with nonzero isospin is the pion. Hence the onset at zero temperature takes place at a critical quark chemical potential $\mu_c$, which equals

\begin{figure}[t]
\centerline{\includegraphics[height=5.5cm]{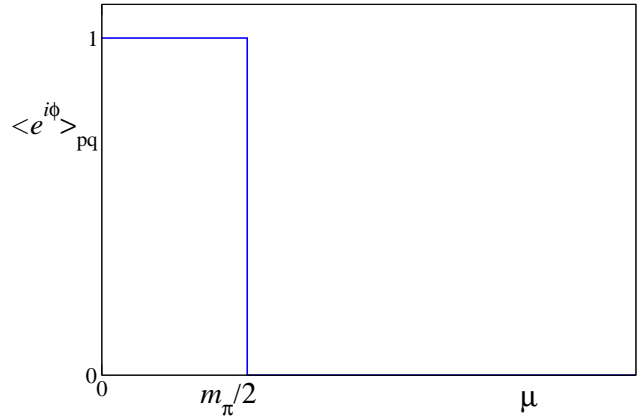}}
\caption{Average phase factor in the thermodynamic limit in the phase-quenched theory at $T=0$.}
\label{fig:av}
\end{figure}

\begin{itemize}
\item for quark chemical potential: $\mu_c=$ [nucleon mass $m_N$ $-$ binding energy]/3 $\quad\Rightarrow\quad$ transition to nuclear matter;
 \item for isospin chemical potential: $\mu_c =$ [pion mass $m_\pi$]/2  $\quad\Rightarrow\quad$ pion condensation.
\end{itemize}
At strictly zero temperature, we find therefore that in the interval $0<\mu<m_\pi/2$ the full and the phase-quenched theories are identical, but no interesting physics is taking place since the thermodynamic quantities are independent of chemical potential.
On the other hand, in the interval $m_\pi/2<\mu\lesssim m_N/3$ strong cancelations are required to cancel the $\mu$ dependence in the full theory. These cancelations are arising from the phase factor, ignored in the phase-quenched theory.
Therefore, in the thermodynamic limit, the average phase factor will be as shown in Fig.~\ref{fig:av}.

We conclude that if the full theory is studied via simulations of the phase-quenched theory, excessive cancelations of the $\mu$ dependence should take place in the region $m_\pi/2<\mu\lesssim m_B/3$. It turns out that this is quite a severe constraint to do this correctly, and many straightforward numerical methods fail this test! As mentioned earlier, this phenomenon has been dubbed the Silver Blaze problem, named after a Sherlock Holmes story (see Ref.\ \cite{Cohen:2003kd}).

In \ref{App:A},
it is shown that the same feature is present in the Bose gas: here the phase-quenched theory is simply a theory with a $\mu^2$-dependent effective mass parameter $m^2_{\rm eff}=m^2-\mu^2$ and the region of severe cancelations (the Silver Blaze region) is given by $0<\mu<m$.

\subsection{Silver Blaze problem and the Dirac operator}

The original formulation of the Silver Blaze problem was given in the context of the eigenvalues of the Dirac operator \cite{Cohen:2003kd}.
Since the weight and therefore configurations in lattice simulations depend on the chemical potential, so should the eigenvalues. Yet, as mentioned several times above,  this $\mu$  dependence should cancel in the expectation value of thermodynamic quantities. In order to achieve this,  it was found that the density of the Dirac eigenvalues has to behave in a highly nontrivial manner 
 \cite{Cohen:2003kd,Osborn:2005ss,Osborn:2008jp}.
We will now briefly describe this.

Let us write the Dirac operator as
\be
M = D+m \qquad \mbox{with}  \qquad D=D\!\!\!\!\slash +\mu\gamma_4.
\ee
The partition function is written as
\be
Z = \int DU\, \det(D+m)e^{-S_{\rm YM}}= \bra \det (D+m)\ket_{\rm YM},
\ee
where the subscript {\scriptsize YM} indicates the average over the gluonic field only (and with slight abuse of notation, the brackets $\bra\cdot\ket_{\rm YM}$ are not normalised).
The determinant is the product of the eigenvalues, 
\be
\det(D+m) = \prod_k \left(\lambda_k+m\right)\qquad\quad\quad D\psi_k=\lambda_k\psi_k.
\ee
Note that since $D$ is not  $\gamma_5$ hermitian at nonzero $\mu$, the eigenvalues not purely real or imaginary, but they are complex in general.

It is customary to look at the chiral condensate, which is expressed as ($\Omega$ denotes again the spacetime volume)
\be
\bra\bar\psi\psi\ket = \frac{1}{\Omega}\frac{\partial \ln Z}{\partial m} = \frac{1}{Z}\left\bra \frac{1}{\Omega}\sum_k \frac{1}{\lambda_k+m}\prod_j(\lambda_j+m)\right\ket_{\rm YM},
\ee
since the derivative with respect to $m$ removes every  factor $\lambda_k+m$ from the determinant once. 
This expression can be written succinctly in terms of the density of eigenvalues, defined as
\bea
\rho(z;\mu)  =&&\hm  \frac{1}{Z}\int DU \, \det(D+m) e^{-S_{\rm YM}} \frac{1}{\Omega}\sum_k\delta^2(z-\lambda_k) 
\nn \\
=&&\hm \frac{1}{Z}\left\bra \det(D+m)  \frac{1}{\Omega}\sum_k\delta^2(z-\lambda_k)  \right\ket_{\rm YM}.
\eea
We can then finally write
\be
\bra\bar\psi\psi\ket = \int d^2z\, \frac{\rho(z;\mu)}{z+m},
\ee
i.e.\ the chiral condensate is given by an integral in the complex plane over the spectral density.

Now, in general $\rho(z;\mu)$ will depend on $\mu$, since the Dirac operator $D$ does. In fact, for every fixed configuration of gauge fields at nonzero $\mu$, there will be explicit $\mu$ dependence. However, once the average over all gauge fields is taken, a.k.a.\ the integral over the spectral density is performed, the $\mu$ dependence should cancel in the region where 
$\mu\lesssim m_B/3$, i.e.\ below onset, in the thermodynamic limit $\Omega\to \infty$.
This is achieved as follows \cite{Osborn:2005ss,Osborn:2008jp}: in the Silver Blaze region, $\rho(z;\mu)$ is a complex function, oscillating with amplitude $e^{\Omega\mu}$ and period $1/\Omega$. Only when all oscillations are correctly integrated, $\mu$  dependence will cancel. This provides the resolution of the Silver Blaze problem, from the viewpoint of the eigenvalues of the Dirac operator. Detailed studies of the Dirac spectral density and the interplay with the sign problem can be found in Refs.\ \cite{Osborn:2005ss,Osborn:2008jp,Splittorff:2007ck,Splittorff:2006fu,Splittorff:2007zh}.

In \ref{App:B} this is worked out in detail in the form of an exercise for QCD in one dimension, in the case of the gauge group U(1).

\section{Phase boundary at small chemical potential}
\label{sec:pb}

The overlap problem is severe at low temperatures, making a lattice study of cold and dense matter prohibitively difficult using standard techniques. However, another relevant question for the QCD phase diagram concerns the thermal transition from the hadronic phase to the quark-gluon plasma when $\mu\sim 0$. Here one may expect the overlap problem to be less severe, since the theories with quark and isospin chemical potential are more alike, and also that approximate methods exploiting the relative smallness of the chemical potential with respect to the temperature can be employed successfully. In this section, we discuss this in some detail.

\begin{figure}[h]
\centerline{ \includegraphics[height=5cm]{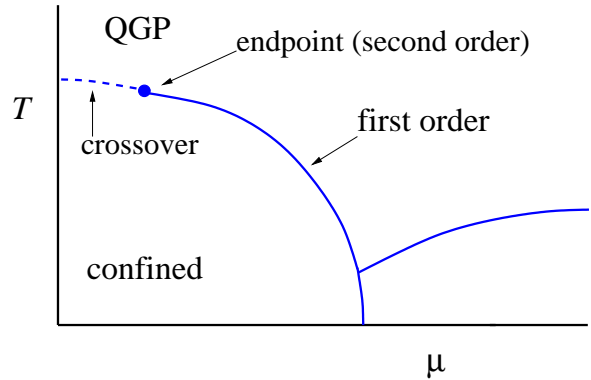}}
\caption{``Standard'' phase diagram.}
\label{fig:pds}
\end{figure}

The aim will be to determine the phase boundary between the confined and deconfined phase at small $\mu$ and, possibly, locate the critical endpoint in the phase diagram, assuming it exists. As a reminder, a sketch of the ``standard'' phase diagram is shown in Fig.\ \ref{fig:pds}. For physical quark masses, the transition at $\mu=0$ is a crossover \cite{Aoki:2006we}, and one may expect this to change into a first-order transition at larger $\mu$. The critical endpoint marks the end of the first-order line \cite{Stephanov:2004wx}.
We note here that the investigation of this part of the phase diagram is very well motivated from an experimental point of view, due to the ongoing and planned heavy-ion collisions at RHIC, the LHC and hopefully at FAIR at GSI.

For small chemical potential, the critical temperature of the phase boundary at nonzero $\mu$ can be written as a series in $\mu/T$ \cite{Allton:2002zi,deForcrand:2002ci}, for instance as
\be
\frac{T_c(\mu)}{T_c(0)} = 1+ \#\left(\frac{\mu}{T_c(0)}\right)^2 +  \#\left(\frac{\mu}{T_c(0)}\right)^4  +\ldots
\ee
Since the partition function is an even function of $\mu$, only even powers of $\mu$ appear. Note that if the transition is a crossover, a unique transition line is not defined and hence the coefficients in this expansion and also $T_c(0)$ may depend on the observable. To avoid too many notational complications, we will refer to $T_c$ as the generic transition temperature, defined in a suitable manner in the case of a crossover.
A considerably more advanced step is to also attempt to determine the location of the critical endpoint from the radius of convergence of the expansion, with knowledge of the first few coefficients only \cite{Gavai:2004sd}.

In the following we discuss several approaches which have been used to determine the phase boundary.

\subsection{Reweighting}

The general strategy in reweighting was already discussed above and is summarised here. The partition function is written as
\be
Z_w = \int DU\, w(U), \qquad\quad\quad w(U) \in \mathbf{C},
\ee
and observables are expressed as
 \be
 \bra O\ket_w = \frac{\int DU\, O(U)w(U)}{\int DU\, w(U)}.
\ee
Let us now introduce a new weight $r(U)$ ($r$ for `reweighting' or `real'), chosen at will, such that 
\be
\bra O\ket_w = \frac{\int DU\, O(U)\frac{w(U)}{r(U)} r(U)}{\int DU\, \frac{w(U)}{r(U)} r(U)} = 
\frac{\bra O \frac{w}{r}\ket_r}{\bra \frac{w}{r}\ket_r}.
\ee
As above, the reweighting factor indicates the severity of the overlap problem,
\be
 \left\bra \frac{w}{r}\right\ket_r = \frac{Z_w}{Z_r}= e^{-\Omega\Delta f}, \qquad\quad\quad
\Delta f = f_w-f_r\geq 0,
\ee
where $\Omega$ denotes again the spacetime volume.

\begin{figure}[t]
\centerline{ 
\includegraphics[height=4.3cm]{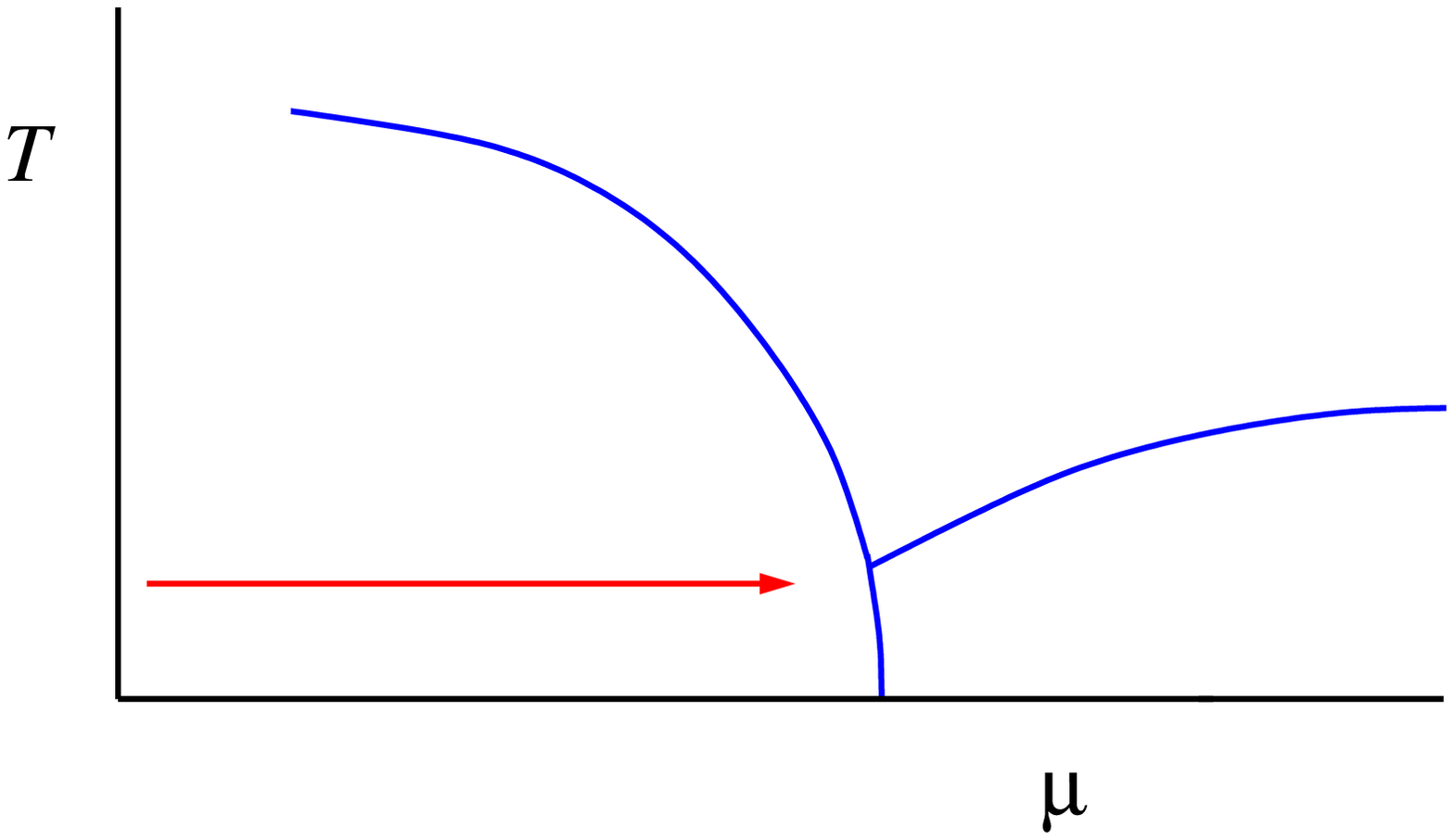} $\qquad$
\includegraphics[height=4.3cm]{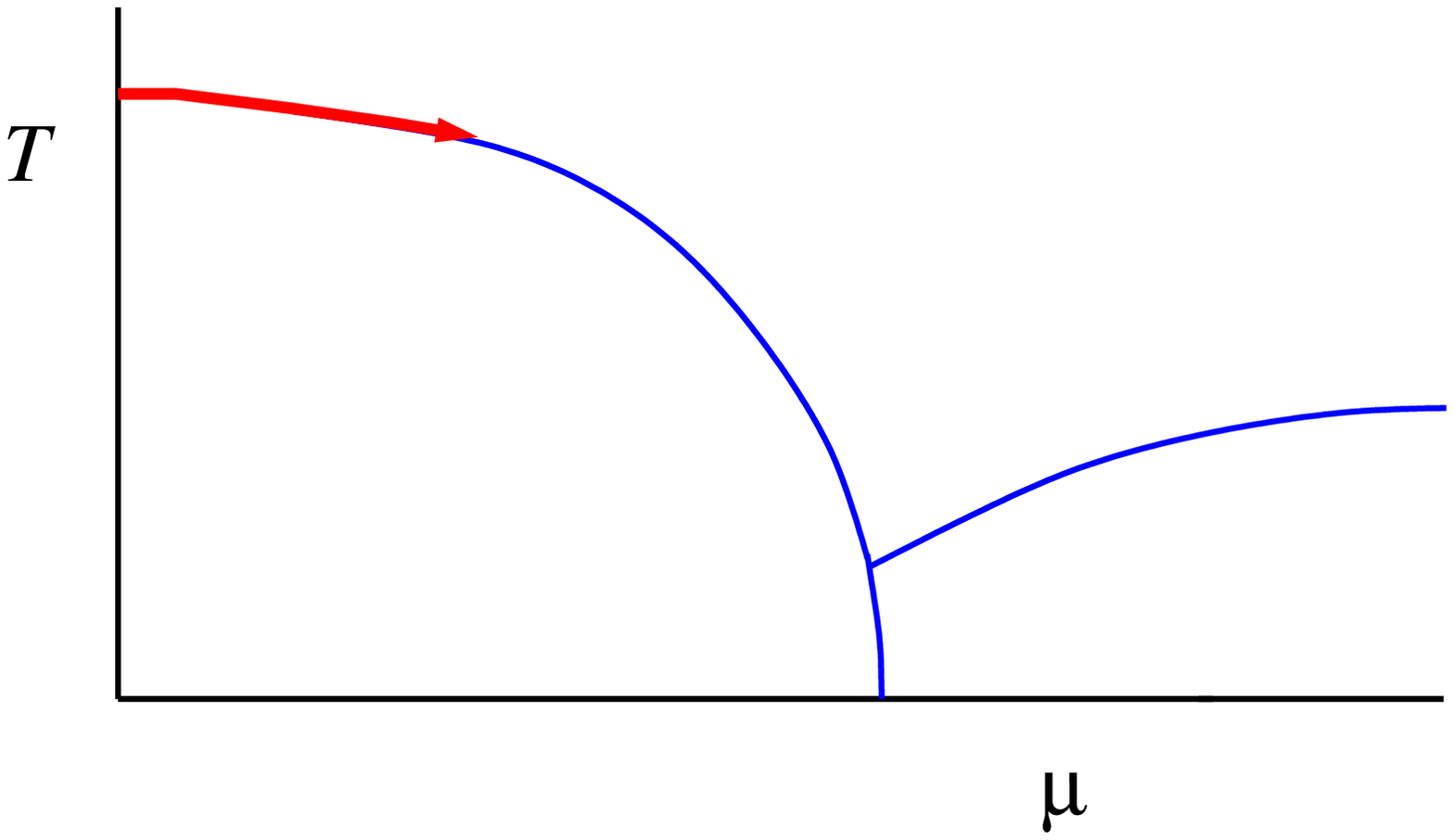}}
\caption{Reweighting at fixed temperature (left) and multiparameter reweighting, aiming at maximising the overlap as well as possible (right). }
\label{fig:rew}
\end{figure}

There is considerable freedom in choosing the new weight $r(U)$, provided that it has the interpretation of a probability weight, such that numerical simulations are possible. Hence one may adapt $r$ to the  problem at hand. Two examples are
\begin{itemize}
\item
Glasgow reweighting \cite{Barbour:1997ej}: work at a fixed temperature (or lattice coupling $\beta$) and  increase $\mu$ from 0, i.e.
\be
\frac{w}{r}\sim \frac{\det M(\mu)}{\det M(0)},
\ee
as illustrated in Fig.\ \ref{fig:rew} (left).
 However, this choice has a severe overlap problem, since the high-density phase is probed with hadronic physics at $\mu=0$. One expects $\Delta f$ to be large and hence the overlap problem will appear already on small volumes. 
\item multi-parameter/overlap preserving reweighting \cite{Fodor:2001pe}: here the temperature (or lattice coupling $\beta$) is adapted as well, see  Fig.\ \ref{fig:rew} (right). Hence 
\be
\frac{w}{r}\sim \frac{\det M(\mu)}{\det M(0)}e^{-\Delta S_{\rm YM}},
\ee
where $\Delta S_{\rm YM}$ is the difference between gauge actions with different gauge couplings.
The main idea here is to attempt to stay on the pseudo-critical line $T_c(\mu)$, hence improving or even ensuring overlap, since both the hadronic phase and the quark-gluon plasma are sampled during the numerical simulation.
This approach has led to a determination of the location of the critical endpoint \cite{Fodor:2004nz}: namely $\mu^q_E=120(13)$ MeV, $T_E = 162(2)$ MeV, see Fig.\ \ref{fig:FK}. This result was obtained using  $N_f=2+1$ quark flavours with physical quark masses on a coarse lattice with $N_\tau=4$ points in the temporal direction. Unfortunately, this method is very expensive to extend to smaller lattice spacing (larger $N_\tau$)  and it has not been repeated. 
A critical analysis can be found in Ref.\ \cite{Splittorff:2006vj}
\end{itemize}

\subsection{Taylor series expansion}

An alternative, and more modest, idea relies on a Taylor series expansion in $\mu/T$ around $\mu=0$. The coefficients in the expansion can be calculated using conventional simulations at $\mu=0$, where the sign problem is absent. This approach continues to be  pursued by several groups \cite{Allton:2002zi,Gavai:2004sd,Allton:2005gk,Kaczmarek:2011zz,Endrodi:2011gv,Borsanyi:2012cr}. A recent review can be found in Ref.\  \cite{Borsanyi:2015axp}.

Let us start again from the grand-canonical ensemble, or pressure, 
\be
p=\frac{T}{V}\ln Z.
\ee
Using that the pressure is an even function of $\mu$, we can write
\be
\Delta p(\mu) \equiv p(\mu) - p(0) 
=  \frac{\mu^2}{2!}\frac{\partial^2p}{\partial\mu^2} \Big|_{\mu=0}+  \frac{\mu^4}{4!}\frac{\partial^4p}{\partial\mu^4} \Big|_{\mu=0} + \ldots,
\ee
or more compactly,
\be
\frac{\Delta p(\mu)}{T^4} = \sum_{n=1}^\infty c_{2n}(T) \left(\frac{\mu}{T}\right)^{2n}.
\ee
 The coefficients $c_{2n}$ are defined at $\mu=0$. Note that other thermodynamic quantities follow immediately, for example the density is given by 
 \be
 \bra n(\mu)\ket =  \frac{\partial p}{\partial\mu}  = 2T^3 \sum_{n=1}^\infty n c_{2n}(T) \left(\frac{\mu}{T}\right)^{2n-1}.
 \ee

 \begin{figure}[t]
\centerline{ 
\includegraphics[height=5.5cm]{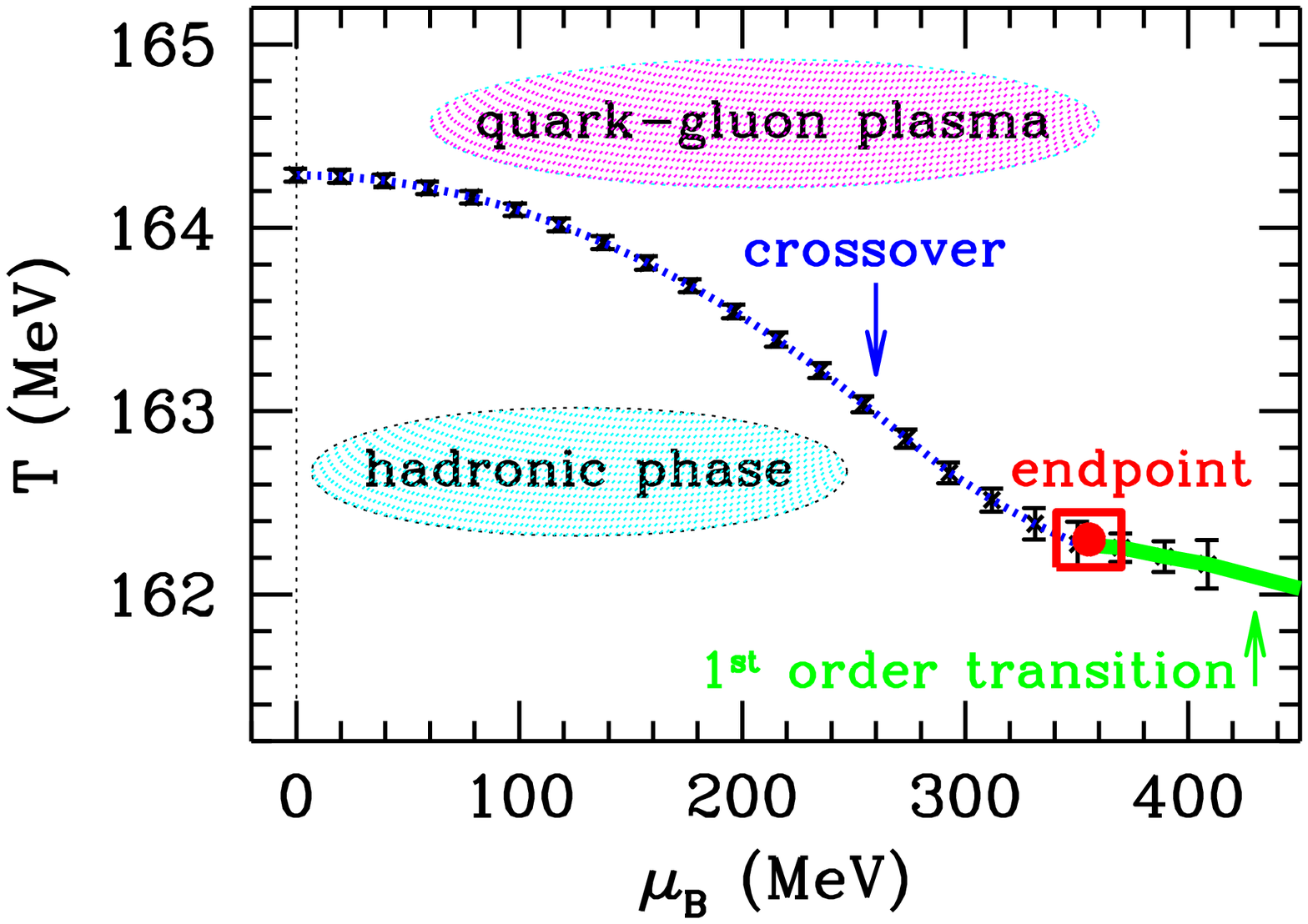}
\includegraphics[height=5.5cm]{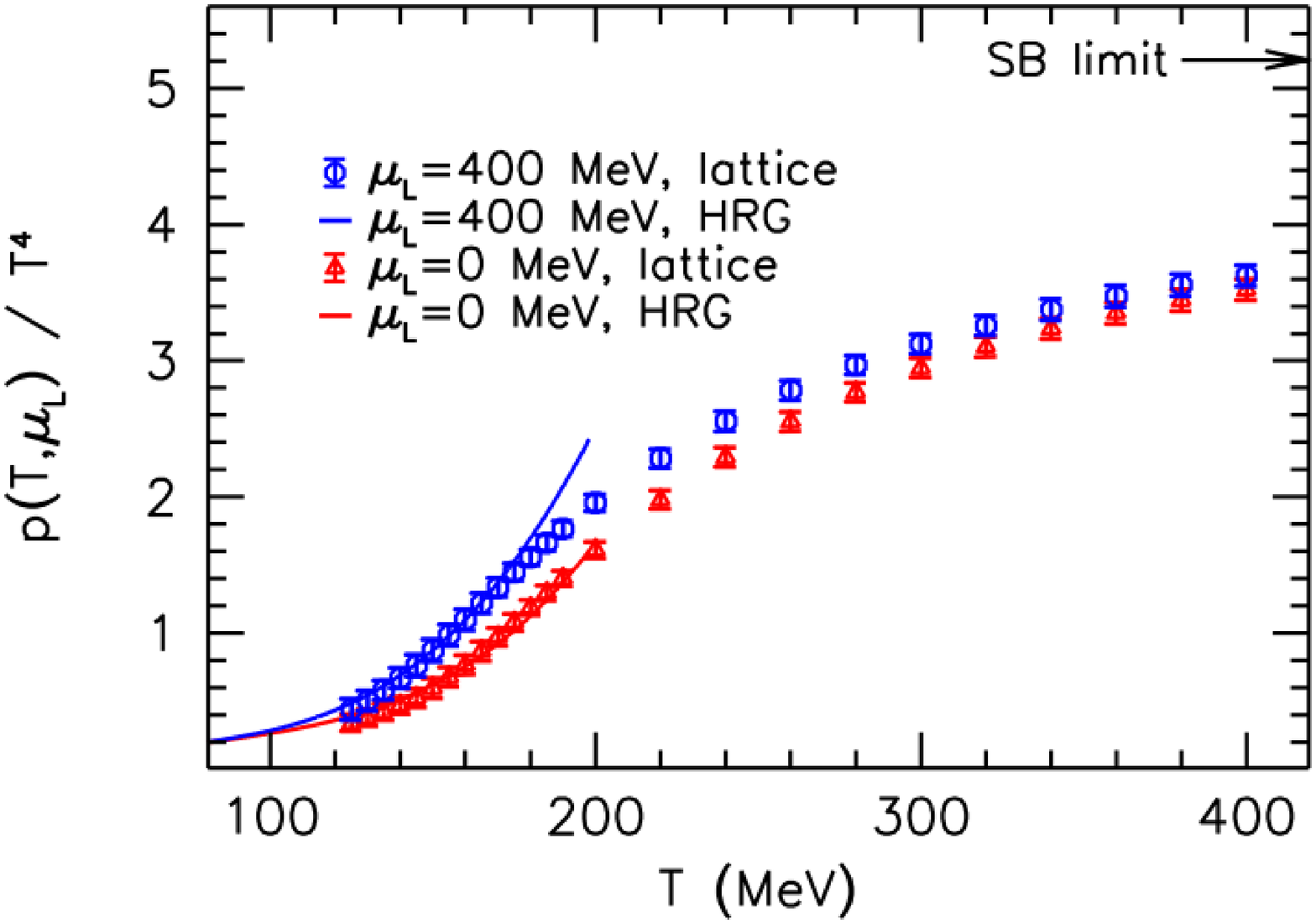}
}
\caption{Left: Location of the critical endpoint using multi-parameter/overlap preserving reweighting, on a lattice with $N_\tau=4$ \cite{Fodor:2004nz}.
Continuum estimate of the pressure as a function of temperature for $\mu_L=0$ and 400 MeV, only including the term up to ${\cal O}(\mu_L^2)$, for $N_f=2+1$ flavours of quarks with physical masses, using an continuum extrapolation \cite{Borsanyi:2012cr}. Here $\mu_L$ refers  to the baryon chemical potential for the two light flavours only.
}
\label{fig:FK}
\end{figure}

 In order to see what it is needed in practice, it is useful to give some explicit expressions. We start from
\be
Z = \int DU\left(\det M\right)^{N_f} e^{-S_{\rm YM}} = \int DU\, e^{-S_{\rm YM}+N_f\ln\det M(\mu)}.
\ee
Differentiation is straightforward and 
\bea
\nn
\frac{\partial\ln Z}{\partial\mu} = &&\hm \left\bra N_f\frac{\partial}{\partial\mu}\ln\det M\right\ket, \\
\frac{\partial^2\ln Z}{\partial\mu^2} = &&\hm 
\left\bra N_f\frac{\partial^2}{\partial\mu^2}\ln\det M\right\ket +
\left\bra \left(N_f\frac{\partial}{\partial\mu}\ln\det M\right)^2\right\ket  -\left\bra N_f\frac{\partial}{\partial\mu}\ln\det M\right\ket^2, 
\eea
etc. Writing $\ln \det M = \Tr\ln M$, these can be expressed as
\bea
\nn
\frac{\partial}{\partial\mu}\ln\det M= &&\hm \Tr M^{-1}\frac{\partial M}{\partial\mu}, \\
\frac{\partial^2}{\partial\mu^2}\ln\det M= &&\hm \Tr M^{-1}\frac{\partial^2 M}{\partial\mu^2} -
 \Tr M^{-1}\frac{\partial M}{\partial\mu} M^{-1}\frac{\partial M}{\partial\mu},
\eea
etc., allowing for an easy diagrammatic interpretation. 
It is straightforward to work out more derivatives, but the number of terms increases rapidly, see e.g.\ Ref.\ \cite{Gavai:2004sd} for terms contributing to quite high order.
  Moreover, there are again cancelations required: the pressure $p$ is an intensive quantity, and hence the coefficients $c_{2n}$ are finite in the thermodynamics limit. However,  the individual contributions may scale differently, as is clear from the explicit expressions above. This situation is familiar from e.g.\ the usual (second-order) susceptibilies $\chi$, but is enhanced at higher order, where the $c_{2n}$'s can be viewed as generalized susceptibilities.

Most current work focuses on going closer to the continuum limit for physical quark masses. An example is given in Fig.\ \ref{fig:FK} (right)  \cite{Borsanyi:2012cr}: plotted is a continuum estimate of the pressure as a function of temperature for two values of $\mu_L$, the baryon chemical potential for the two light flavours. Note that only the
 ${\cal O}(\mu_L^2)$ contribution is included.
 
 \begin{figure}[t]
\centerline{ \includegraphics[height=5.5cm]{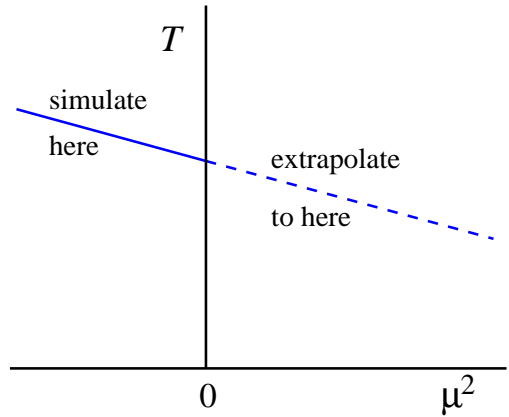}}
\caption{Phase boundary around $\mu^2=0$ in the $T-\mu^2$ plane.}
\label{fig:ana}
\end{figure}

\subsection{Imaginary $\mu$}

As shown in Sec.\ \ref{sec:qs} the fermion matrix satisfies the property
\be
[\det M(\mu)]^* = \det M(-\mu^*).
\ee
Hence its determinant is real if the chemical potential is chosen to be purely imaginary, $\mu=i\mu_{\rm I}$. One is then able to perform ordinary simulations employing importance sampling. In particular, one may obtain the transition line $T_c(\mu_{\rm I})$ at imaginary $\mu$. Treating $\mu$ as a complex parameter the transition line for real $\mu$ can then be obtained by analytical continuation. Since $T_c$ is even in $\mu$, this is particularly straightforward and amounts to $+\mu^2_{\rm I} \to -\mu^2$. Hence one may follow these steps, as illustrated in Fig.\ \ref{fig:ana}:
\begin{enumerate}
\item
determine the phase boundary  at $\mu^2<0$;
\item parametrise and fit $T_c(-\mu^2)$;
\item obtain the phase boundary at $\mu^2>0$.
\end{enumerate}

This approach has been carried out during the past decade in great detail; an incomplete list includes Refs.\ 
\cite{Lombardo:1999cz,deForcrand:2002ci,D'Elia:2002gd,deForcrand:2003hx,deForcrand:2008vr,Cea:2012ev,Bonati:2014kpa}.
 However, it turns out that at imaginary $\mu$ QCD has quite an  intricate phase structure and that the topic is much richer than just for applying analytical continuation around $\mu^2\sim 0$. Hence we will discuss QCD at imaginary $\mu$ in more detail in the next section.

\subsection{Summary}

\begin{figure}[t]
\begin{center}
\begin{minipage}{6.5cm}
\mbox{}\\
\vspace*{1.5cm}

imaginary $\mu$\\
2 parameter imag.\ $\mu$\\
double reweighting (Lee-Yang zeroes)\\
double reweighting (susceptibilities)\\
canonical
%\end{flushright}}
\end{minipage}$\qquad$
\begin{minipage}{5cm}
\vspace*{-3cm}
\centerline{ \includegraphics[height=9cm]{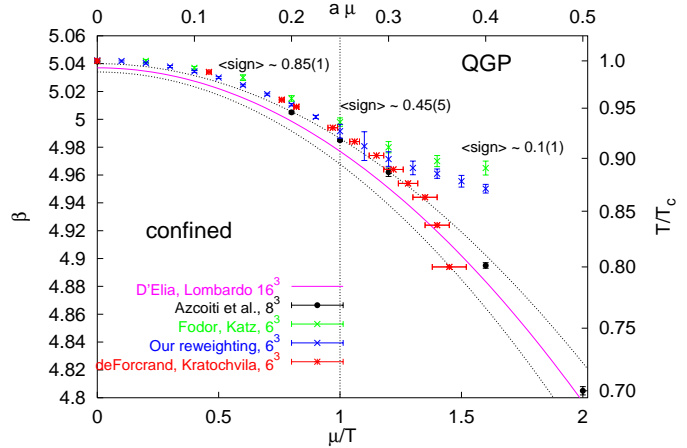}}
\end{minipage}
\begin{minipage}{3cm}\mbox{}\end{minipage}
%\vspace*{-0.3cm}
\caption{Phase boundary in the plane of gauge coupling and chemical potential in lattice units, or, in physical units, $\mu/T$ and $T/T_c$, comparing several methods, for four flavours of staggered quarks on a lattice with  $N_\tau=4$ sites in the temporal direction  \cite{deForcrand:2010ys}.
}
\label{fig:PdF}
\end{center}
\end{figure}

A compilation of results for the phase boundary using various methods is presented in Fig.\ \ref{fig:PdF}. This plot from 2009 \cite{deForcrand:2010ys} is by now already rather old, but it shows the essential findings. Good agreement exists between the various methods as long as $\mu/T\lesssim 1$, for which the average sign is distinctly different from zero, at least on the small spatial volume sizes and fixed $N_\tau=4$ considered.
However, as the chemical potential is increased, the average sign becomes zero within the error and the results from the various approaches start to deviate. Which result is correct, if any, cannot be concluded. 
 Hence the sign problem is preventing further progress.

In more recent years the attention has shifted to the determination of the lowest-order coefficients in the expansion to higher precision than before, i.e.\ for physical quark masses and closer to the continuum limit.
For instance, Ref.\ \cite{D'Elia:2015rwa} gives a summary for results for the second-order coefficient $\kappa$ in the expansion
\be
 \frac{T_c(\mu_B)}{T_c} = 1-\kappa \left( \frac{\mu_B}{T_c}\right)^2 + {\cal O}\left(\mu_B^4\right),
 \ee
and finds that $0.007 \lesssim \kappa \lesssim 0.018$, depending on the method used. Most results have been obtained  still away from the continuum limit and hence it is expected that a unique answer will emerge eventually.
The state-of-the-art has recently been summarised in Ref.\ \cite{Borsanyi:2015axp}.

Possibilities for the critical endpoint will discussed again at the end of the next section.

\section{Imaginary chemical potential}
\label{sec:imag}

QCD at imaginary chemical potential is a much richer topic than we have seen so far.
It has an intricate phase structure due to the following two reasons:
\begin{enumerate}
\item the interplay of chemical potential and centre symmetry;
\item  the sensitivity of the thermal transition to the masses of the three light quarks ($u,d,s$).
\end{enumerate}   
In this section, we will discuss this in some detail, starting from  the quark mass dependence of the thermal transition, summarised in the so-called Columbia plot. We then discuss centre symmetry in the pure SU(3) gauge theory and with the addition of quarks, and finally extend the Columbia plot to three dimensions, by including the chemical potential. This section is based on a series of papers, see e.g.\ Refs.\ \cite{Roberge:1986mm,deForcrand:2002ci,D'Elia:2002gd,deForcrand:2008vr,D'Elia:2009qz} and especially Ref.\ \cite{deForcrand:2010he}.

\subsection{Quark mass dependence of the thermal transition }

\begin{figure}[t]
\centerline{ \includegraphics[height=6cm]{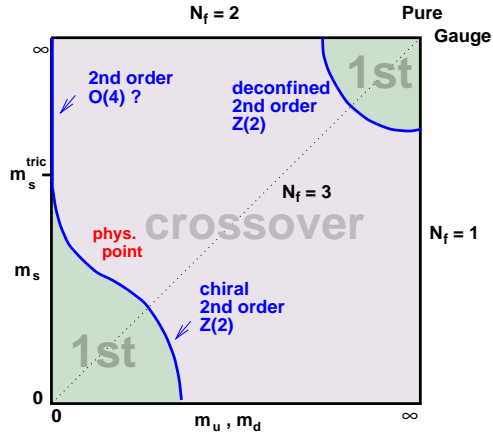}}
\caption{Columbia plot: quark mass dependence of thermal transition.}
\label{fig:Columbia}
\end{figure}

Characteristics of the thermal transition are very sensitive to the masses of the three lightest quarks, $m_q=m_{u,d,s}$. This is summarised in the Columbia plot, shown in Fig.\ \ref{fig:Columbia}, through which we walk now. The horizontal axis indicates the (degenerate) $u$ and $d$ quark masses, and the vertical axis the $s$ quark mass, ranging from 0 to $\infty$. 
Of course, changing the quark masses is  not possible in Nature, but it is possible and very useful in theoretical computations, as it offers additional insight in the phase structure of the strong interaction.

Let us start in the top right corner: $m_{q}=\infty$. Since the quarks decouple, this corner corresponds to the pure SU(3) gauge theory. As will be discussed shortly, the confinement/deconfinement transition can be defined here with the help of $\mathbb{Z}_3$ centre symmetry, which is unbroken at low temperature and broken spontaneously at high temperature. The Polyakov loop acts as the order parameter and the transition is first order. In the presence of quarks with a finite mass (rather than infinite), the centre symmetry is broken explicitly and the Polyakov loop is no longer a true order parameter. Yet the deconfinement transition remains first order for large, but not infinite, quark masses. 

Next we consider the bottom left corner. Here the quarks are massless, $m_{q}=0$, and hence chiral symmetry can be used to define the transition, with the  chiral condensate as the order parameter. Again the transition is first order, and remains first order for small, but nonzero quark massess.
Increasing the quark masses more, or reducing them from infinity, the transition becomes a crossover with no change in symmetry properties (both chiral and centre symmetry are broken explicitly). The lines where the change from first order to crossover happens are lines of second-order phase transitions, as indicated in the figure.
The physical point, with the quark masses as in Nature, is in the crossover region \cite{Aoki:2006we}.

Finally, there are some special choices of quark masses: the diagonal line $N_f=3$ corresponds to three degenerate flavours; the line $N_f=2$ corresponds to two degenerate flavours ($m_s=\infty$); and the line $N_f=1$ to one-flavour QCD ($m_{u,d}=\infty$). The vertical line at $m_{u,d,}=0$ indicates two massless flavours and a massive $s$ quark. The phase structure at larger $m_s$ is still under investigation
\cite{Bonati:2014kpa}: if the transition turns second order at large $m_s$, the first and the second-order transition meet at a tricritical point, indicated with $m_s^{\rm tric}$.

\subsection{Pure gauge: centre symmetry}

In SU($N$) gauge theories without quarks (or with infinitely heavy quarks), there is an exact global symmetry \cite{'tHooft:1977hy}, which is unbroken at low and spontaneously broken at high temperature (we consider $N=3$, but write $N$ for the number of colours). Let us multiply each temporal link in a fixed time slice with a phase factor $z^k$,
\be
U_{4}(\tau,\xv) \quad\to\quad  z^kU_{4}(\tau,\xv),
\qquad\qquad 
z^k= e^{2\pi i k/N} \quad\quad (k=0,\ldots, N-1).
\ee
These phase factors are elements of the centre of SU($N$), i.e.\ they commute with every element of the group: $z^k\id\in \mathbb{Z}_N$, with 
$\det(z^k\id)=1$. 
This centre transformation leaves the action and the path integral measure invariant and is hence a symmetry of the pure gauge theory. Note that it is not a gauge transformation.
 
\begin{figure}[h]
\centerline{ \includegraphics[height=3cm]{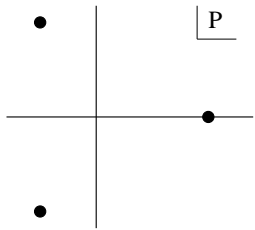}}
\caption{Equivalent vacua in SU(3) gauge theory, in the case of broken centre symmetry $\mathbb{Z}_3$.}
\label{fig:z3}
\end{figure}

The Polyakov loop, the traced product of all links at a spatial site $\xv$ in the temporal direction, 
\be
P(\xv) = \frac{1}{N}\tr \prod_{\tau=0}^{N_\tau-1} U_4(\tau,\xv),
\ee
transforms under this multiplication as 
\be
P(\xv) \to z^k P(\xv).
\ee
Hence if $\bra P\ket =0$, the Polyakov loop expectation value remains zero under a centre transformation and centre symmetry is unbroken. However, 
if $\bra P\ket \neq 0$, a centre transformation will change the expectation value,
\be
 \bra P\ket \to  z\bra P\ket \to  z^2 \bra P\ket \to \ldots \to z^{N-1} \bra P\ket, 
 \ee
 and centre symmetry is broken. Note that the perturbative vacuum corresponds to $U_4=\id$ ($A_4=0$) and therefore $\bra P\ket=1$. Hence centre symmetry is broken perturbatively and there are in fact $N$ equivalent vacua. For $N=3$, this is illustrated in Fig.\ \ref{fig:z3} ($z=1,e^{\pm2\pi i/3}$). 
  Since perturbation theory is relevant at high temperature, we may already expect that at high temperature the centre symmetry is (spontaneously) broken.

This conclusion can be better understood by interpreting the Polyakov loop as the worldline of a massive, i.e.\ static, quark, and the conjugate Polyakov loop, 
\be
P^\dagger(\xv) = \frac{1}{N}\tr \prod_{\tau=N_\tau-1}^0 U_4^\dagger(\tau,\xv),
\ee
as the worldline of a massive anti-quark \cite{Polyakov:1978vu,Susskind:1979up}. Then the free energy $F_{q\bar q}(r)$ for a static quark/anti-quark pair, separated by a distance $r$, is given by
\be
\bra P(\xv) P^\dagger(\yv)\ket = e^{-F_{q\bar q}(r)/T}, \qquad\quad\quad r=|\xv-\yv|.
\ee
At large separation, only the disconnected part of the expectation value survives and
\be
\lim_{r\to\infty} \bra P(\xv) P^\dagger(\yv)\ket = \bra P(\xv) \ket\bra P^\dagger(\yv)\ket = \left| \bra P\ket \right|^2. 
\ee
When there is confinement, the free energy grows with the distance since quarks cannot be separated (there is no string breaking in the pure gauge theory), while in the absence of confinement, the free energy remains finite. Hence we find
\begin{itemize}
\item confined phase: $\;\;\;\;\quad F_{q\bar q}(\infty)\to\infty \qquad\Rightarrow \qquad\bra P\ket =0$,
\item deconfined phase: $\quad F_{q\bar q}(\infty)$ finite $\qquad\Rightarrow \qquad\bra P\ket \neq 0$.
\end{itemize}
The Polyakov loop acts therefore as an order parameter for (de)confinement and centre symmetry is broken in the deconfined phase.

\subsection{Centre symmetry with quarks}

In the presence of quarks, centre symmetry is broken explicitly: for instance the term $\bar\psi_x U_4 \psi_{x+4}$ in the action, see  Eq.\ (\ref{eq:action}), is not invariant under the centre transformation. While in the pure gauge theory, all $\mathbb{Z}(N)$ vacua are equivalent, with quarks the trivial vacuum is preferred, $\bra P\ket\sim 1$. The quarks act as an external symmetry breaking field, choosing a preferred direction, in the same way an external magnetic field acts in the Ising model or a quark mass in the case of chiral symmetry.

However, in the presence of an imaginary chemical potential, multiplying the temporal links as $U_4\to e^{i\mu_{\rm I}} U_4$, something nontrivial happens, since the centre transformation can be undone by a shift in $\mu_{\rm I}$. To see this, let us move all the $\mu_{\rm I}$ dependence to the final time slice, which is possible via a field redefinition, see Eq.\ (\ref{eq:redef}). The chemical potential now appears as $e^{i\mu_{\rm I}/T}$.
If  a $\mathbb{Z}(N)$ transformation is performed on the final time slice, the following combination appears,
\be
z^k e^{i\mu_{\rm I}/T} = \exp i \left( \frac{\mu_{\rm I}}{T} + \frac{2\pi k}{N}\right).
\ee
Hence we note that the centre transformation can be undone by a shift in $\mu_{\rm I}$: this leads to a new symmetry, often called Roberge-Weiss symmetry \cite{Roberge:1986mm},
\be
Z\left( \frac{\mu}{T}\right) = 
Z\left( \frac{\mu}{T}+ \frac{2\pi ik}{N}\right).
\ee
Note that $Z(\mu)=Z(-\mu)$ still holds as well.
Hence the partition function and the phase structure are periodic in the $\mu_{\rm I}$ direction with period $2\pi T/N$ and the range of $\mu_{\rm I}/T$ is limited by $\pi/N$, starting at $\mu_{\rm I}=0$. 

\begin{figure}[t]
\centerline{ 
\includegraphics[height=6cm]{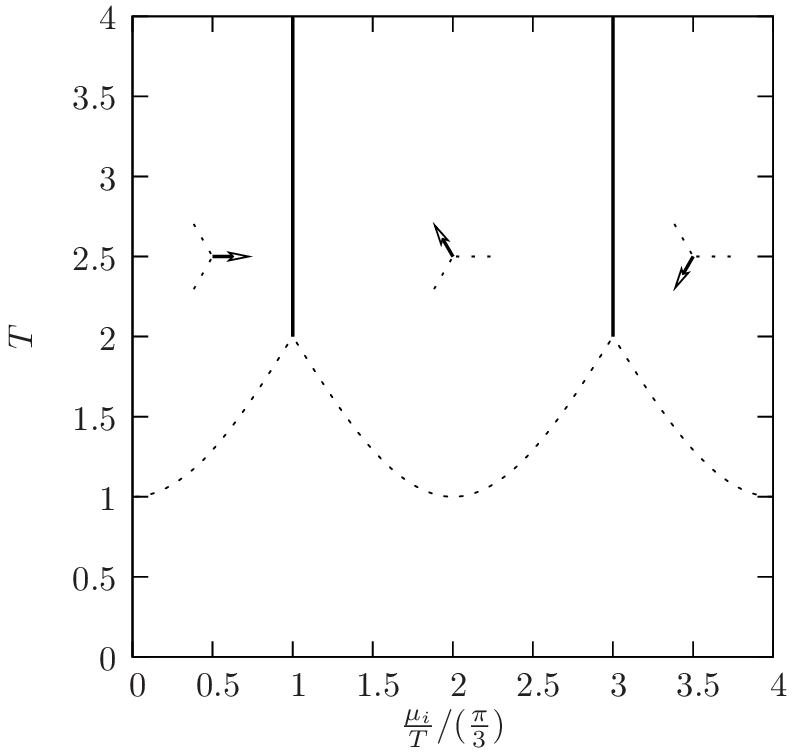} $\qquad$
\includegraphics[height=5.5cm]{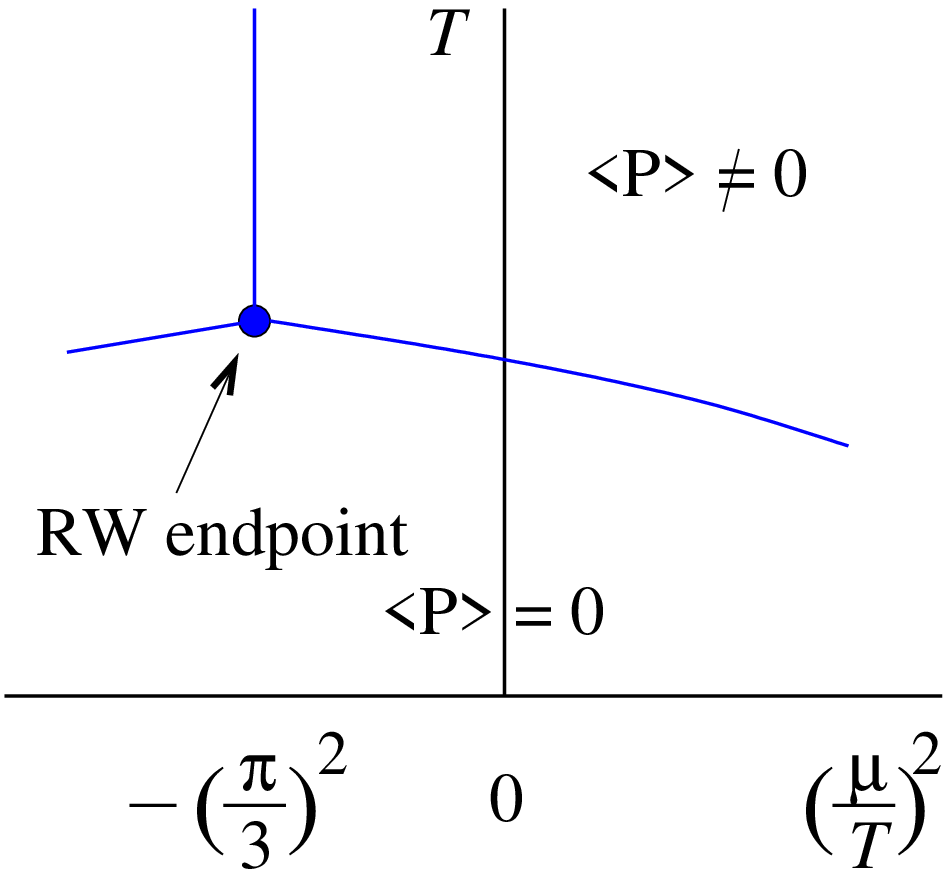}}
\caption{Phase structure in the $\mu_{\rm I}-T$ plane \cite{deForcrand:2010he}
 (left) and the $\mu^2-T$ plane (right). See main text for more details.}
\label{fig:RW}
\end{figure}

Another way to interpret the Roberge-Weiss symmetry is to note that an increase in $\mu_{\rm I}$ is equivalent to a centre transformation.
However, since the Polyakov loop is not invariant under the latter, the choice of preferred vacuum will change as $\mu_{\rm I}$ is increased. 
When $\bra P\ket\neq 0$, the Polyakov loop expectation value cycles through the $N$ different possibilities: the preferred vacuum is given by the
\begin{itemize}
\item trivial vacuum  $\quad\;\;\bra P\ket \sim 1\quad$  at $\quad\mu_{\rm I}/T\sim 0$;
\item rotated vacuum $\quad\bra P\ket \sim z\quad$  at $\quad\mu_{\rm I}/T\sim 2\pi/N$;
\item rotated vacuum $\quad\bra P\ket \sim z^2\quad\!\!$ at $\quad\mu_{\rm I}/T\sim4\pi/N$;
 \end{itemize}
 etc. 
 In the confined phase $\bra P\ket=0$ and this observation is not relevant.  However, in the  deconfined phase,  $\bra P\ket\neq 0$, and the direction of symmetry breaking changes. This is illustrated in Fig.\ \ref{fig:RW} (left) by the symbols with the three little arrows.
The remarkable consequence of this is that exactly at the boundaries, given by $\mu_{\rm I}/T = (2r+1)\pi/N$ ($r=0,1,2,\ldots$), we find again a proper first-order phase transition, with the 
Polyakov loop as order parameter, even in the presence of quarks!

To continue, let us now include the thermal transition line in the phase diagram. Recall that for real $\mu$ we have written
\be
\frac{T_c(\mu)}{T_c} =  1 - \#\left( \frac{\mu}{T_c}\right)^2 +\ldots, 
\qquad\qquad \# > 0.
\ee
Hence this line increases quadratically with $\mu_{\rm I}$, as indicated with the dotted line in Fig.\ \ref{fig:RW} (left). It is natural to connect the thermal transition line with the vertical {\em Roberge-Weiss line} at $\mu_{\rm I}/T=\pi/3$. The point where the lines meet is known as the {\em Roberge-Weiss endpoint}. For larger $\mu_{\rm I}$, the phase structure is determined by the periodicity.

To combine the findings for real and imaginary chemical potential in one diagram, we show the resulting phase structure in the $\mu^2-T$ plane in Fig.\  \ref{fig:RW} (right). The thermal transition line decreases linearly in $\mu^2$ around $\mu^2\sim 0$ and connects to the Roberge-Weiss endpoint on the left.

\begin{figure}[t]
\centerline{
 \includegraphics[height=4cm]{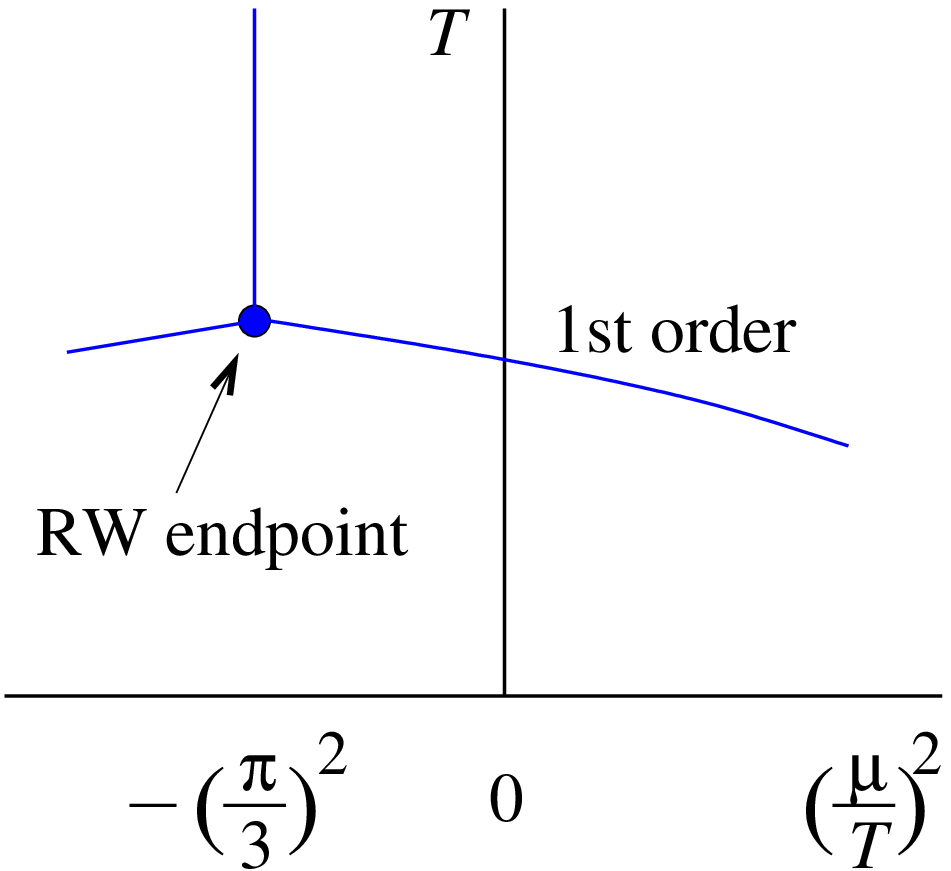} $\quad$
 \includegraphics[height=4cm]{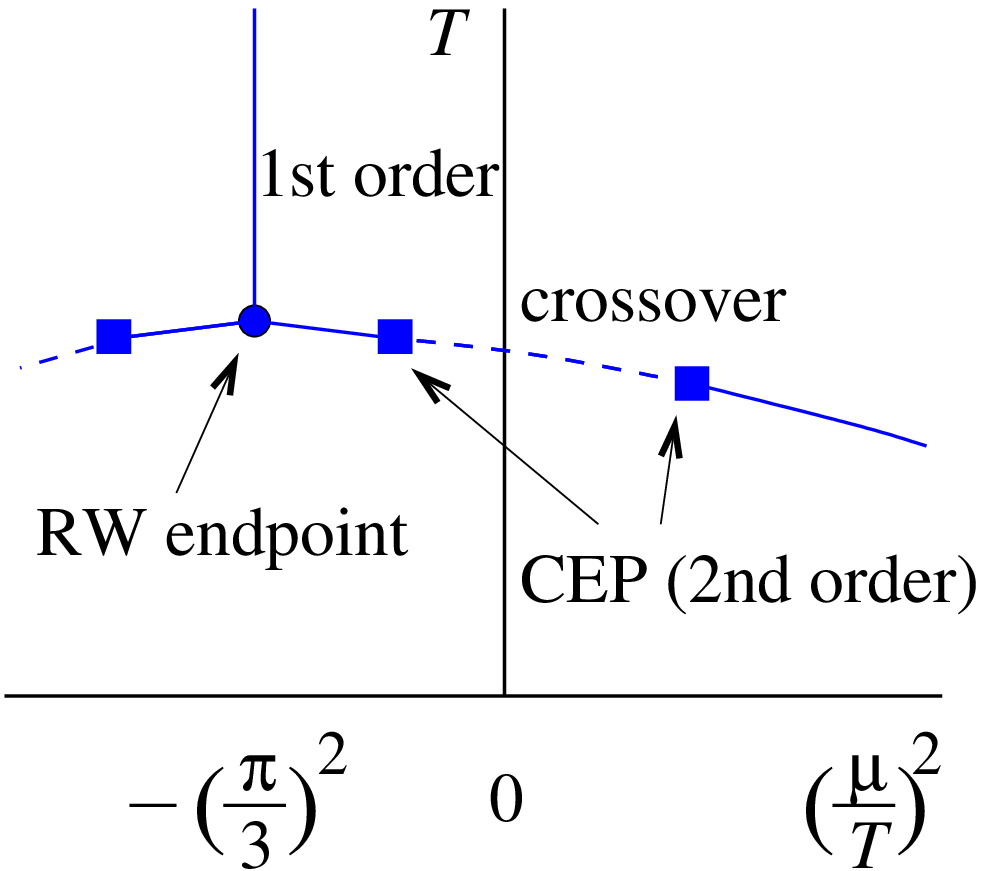} $\quad$
 \includegraphics[height=4cm]{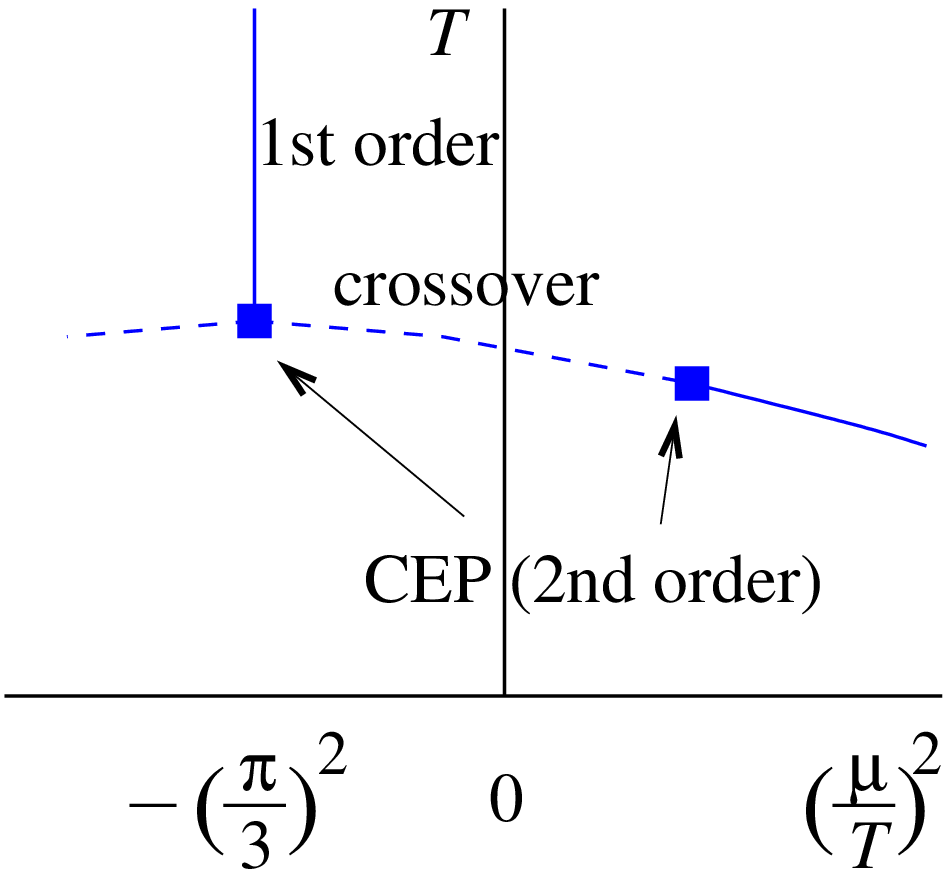}
}
\caption{Phase structure in the $\mu^2-T$ plane for $N_f=3$ degenerate quarks, for very heavy or very light quarks (left), for quarks with an intermediate mass (middle), and for quark masses changed even more (right). The Roberge-Weiss endpoint is a triple point in the first two cases and a critical endpoint in the latter.  }
\label{fig:RW3}
\end{figure}

We saw from the Columbia plot that details of the phase structure depend strongly on the quark masses. At $\mu=0$ the transition is first order for very light or very heavy quarks, and a crossover for intermediate quark masses. This structure extends to nonzero chemical potential as follows (for definiteness we consider the case $N_f=3$):
\begin{itemize}
\item heavy or light quarks:  the first-order transition remains first order for all imaginary $\mu$. At the Roberge-Weiss endpoint, three first-order lines come together, making it triple point. This is illustrated in Fig.\ \ref{fig:RW3} (left);
\item quarks with intermediate mass: the crossover at $\mu^2=0$ turns into a first-order transition at some value of $\mu_{\rm I}$ and possibly also at some value of real $\mu$. The point(s) where this occurs are second-order critical endpoints (CEP). The Roberge-Weiss point is still a triple point, see Fig.\ \ref{fig:RW3} (middle);
\item adapting the quark mass even more: the CEP at imaginary $\mu$ coincides with the Roberge-Weiss point. The transition is a crossover for all values of $\mu_{\rm I}$. 
There might also still be a CEP for real $\mu$, see Fig.\ \ref{fig:RW3} (right).
\end{itemize}

\begin{figure}[t]
 \centerline{ 
\includegraphics[height=5.1cm]{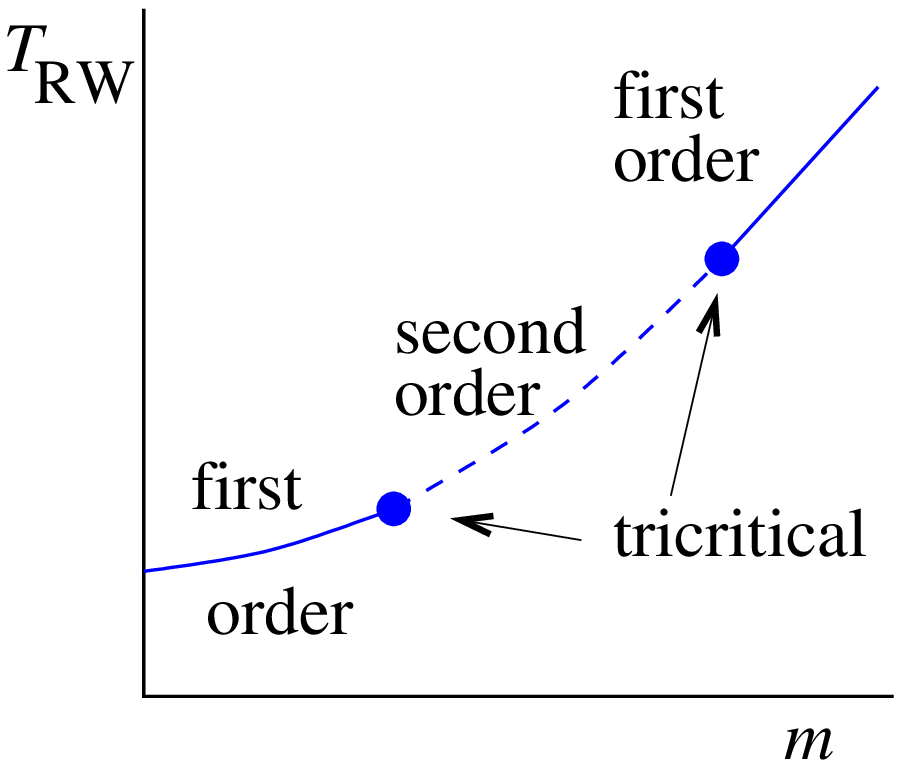}
$\qquad$
\includegraphics[height=5.1cm]{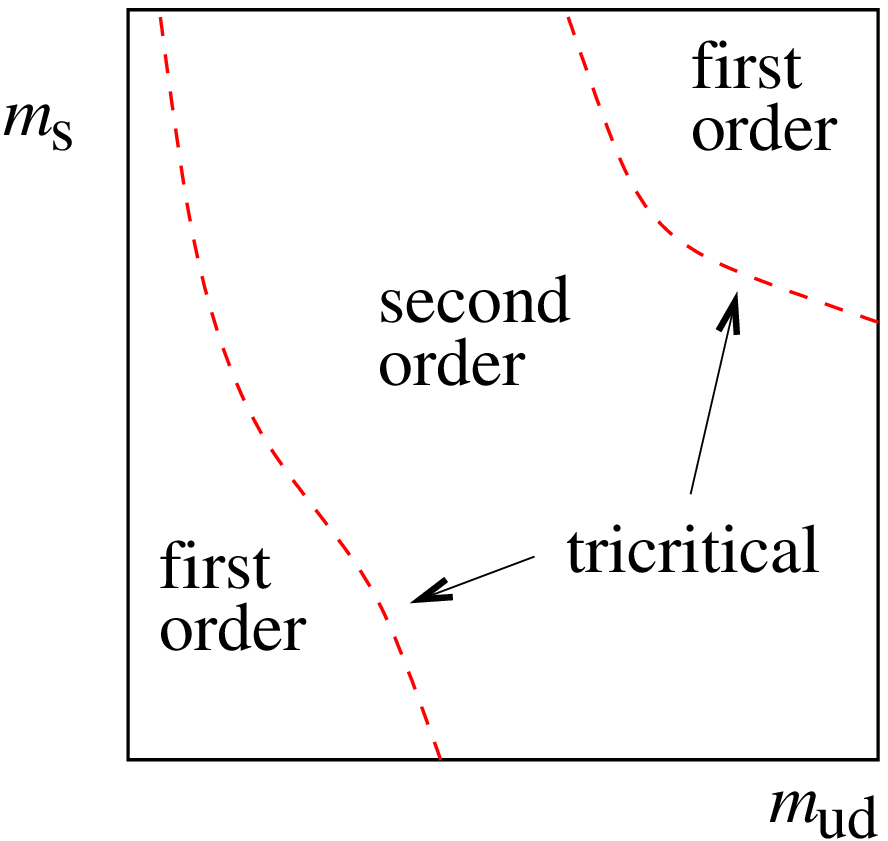} 
 }
\caption{Left: Quark mass dependence of the temperature of the Roberge-Weiss endpoint, $T_{\rm RW}$, for $N_f=3$. Right: equivalent of the Columbia plot at $\mu_{\rm I} = (\pi/3)T$.}
\label{fig:RW4}
\end{figure}

We find therefore that the Roberge-Weiss endpoint is either a first-order triple point, where three first-order lines come together (for heavy and light quarks), or a second-order critical endpoint (for intermediate masses). Note that the temperature of the Roberge-Weiss endpoint depends on the quark mass as well: it increases with quark mass, just as the critical temperature at $\mu=0$ increases with quark mass.
This leads to the result shown in Fig.\ \ref{fig:RW4} (left): the critical temperature $T_{\rm RW}$ as a function of the quark mass, for $N_f=3$. Since the Roberge-Weiss point is second order for intermediate quark masses and first order for larger and smaller masses, there are two tricritical points on this diagram, namely where the first and second-order lines meet.

\subsection{Three-dimensional Columbia plot}

We are now in a position to extend the Columbia plot by including the chemical potential. 
Recall that at $\mu=0$ this plot indicates the order  (first, second or crossover) of the thermal  transition.
First let us consider the case of $\mu_{\rm I}/T= \pi/3$. As argued above, the transition takes place at $T_{\rm RW}$ and it is either first order or second order, depending on the quark masses. Hence a conjectured Columbia plot at $\mu_{\rm I}/T=\pi/3$ is as shown in 
Fig.\ \ref{fig:RW4} (right). We remind the reader that the entire plot is critical and that the boundaries where the first- and second-order transitions meet are tricritial. This should be compared with the Columbia plot at $\mu=0$, where the central region indicates a crossover and the boundaries are second-order lines. Note that Fig.\ \ref{fig:RW4} (left) is  the $N_f=3$ (diagonal) cut through the plot on the right.
The tricritical lines can be determined numerically by varying the quark masses, since there is no sign problem. This amounts to a detailed study of the properties of the Roberge-Weiss endpoint \cite{D'Elia:2009qz,deForcrand:2010he}.

\begin{figure}[t]
\centerline{ 
\includegraphics[height=5.cm]{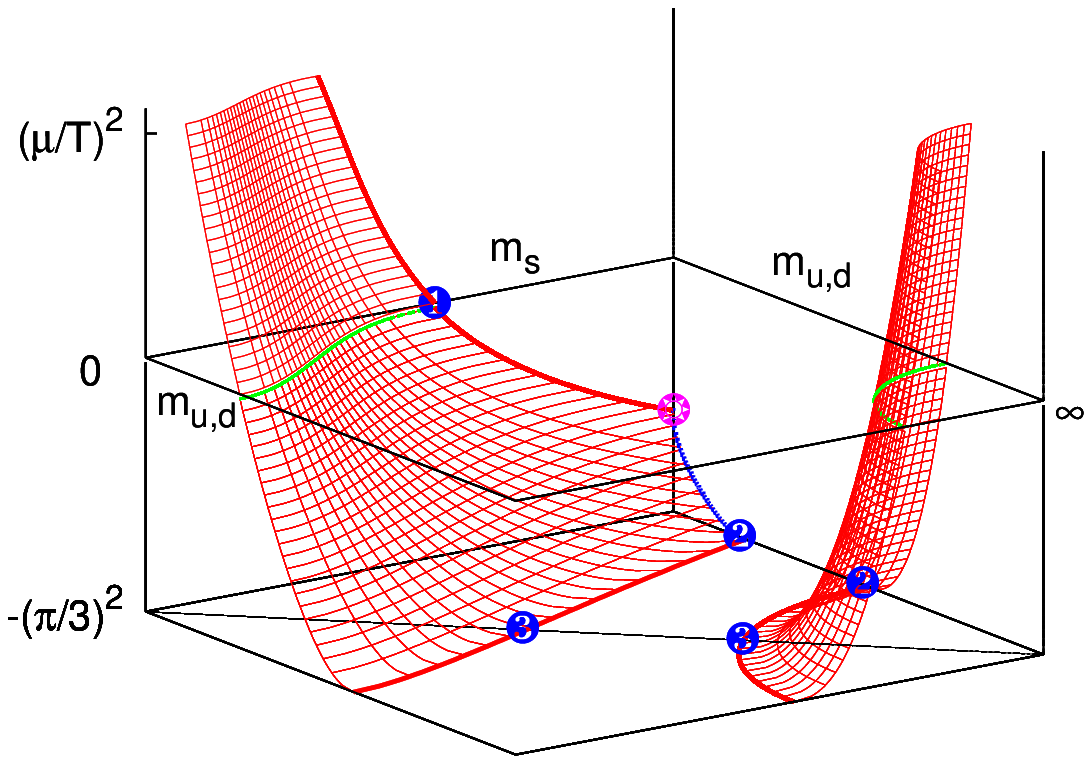}
\includegraphics[height=7.cm]{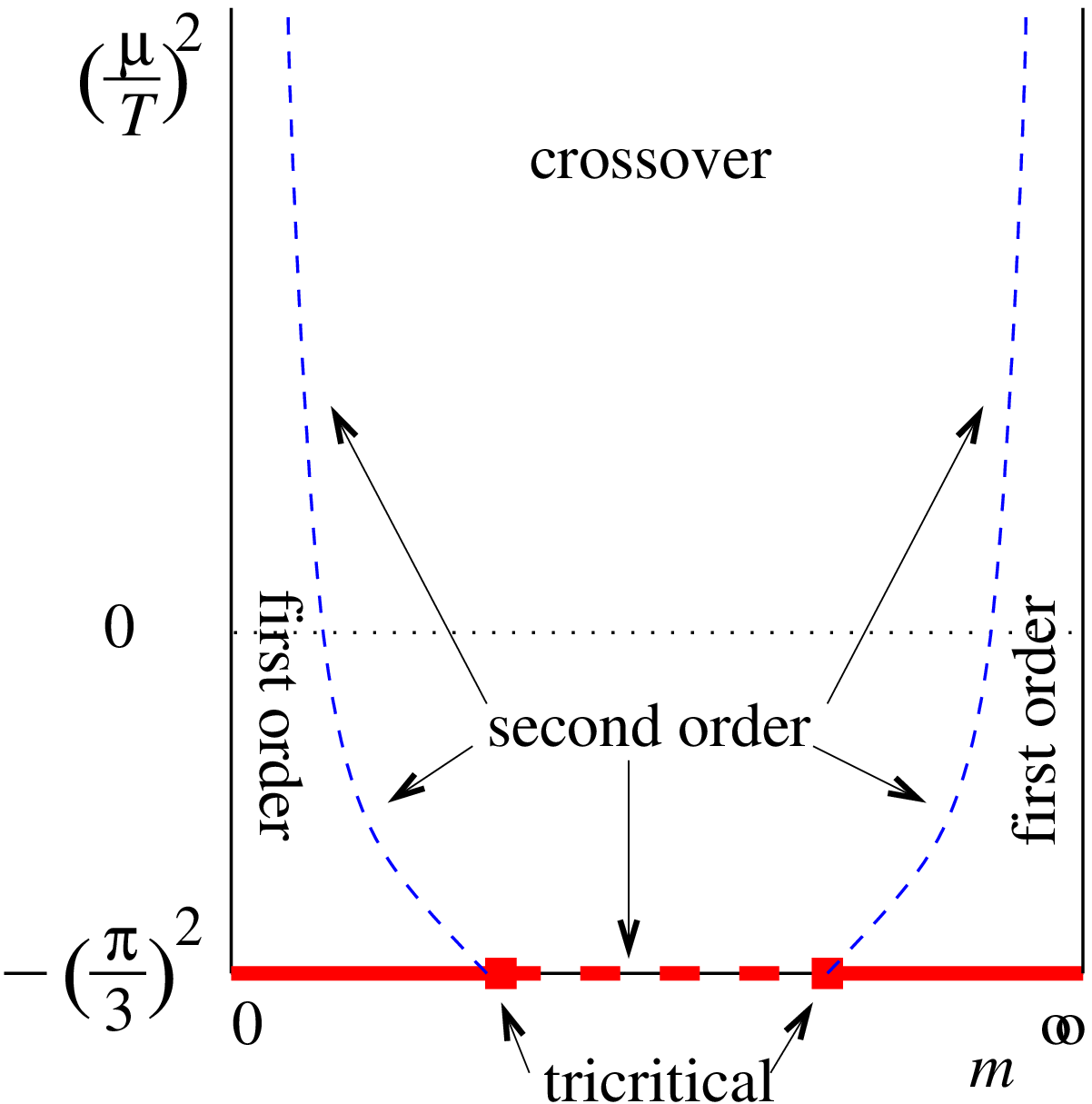}
}
\caption{Columbia plot extended with chemical potential $(\mu/T)^2$ in the vertical direction, for nondegenerate quark masses $m_{u,d}$ and $m_s$ \cite{Bonati:2012pe}
(left) and for $N_f=3$, with $m_{u,d,s}=m$ (right).  }
\label{fig:col2}
\end{figure}

Finally, we can properly extend the Columbia plot with the chemical potential as the third direction. The obvious choice for coordinate is $\mu^2/T^2$, with
\be
 -(\pi/3)^2 \leq  (\mu/T)^2 < \infty,
\ee
where the lower boundary comes from the Roberge-Weiss periodicity. The result is shown in Fig.\ \ref{fig:col2} (left). The red fishnets indicate second-order surfaces, inside of which the transition is a crossover, while near the $m_q=0$ and $m_q\to \infty$ axes, the transition is  first order. The plane $\mu=0$ is the original Columbia plot, while the plane $(\mu/T)^2 =-(\pi/3)^2$ was already shown in Fig.\ \ref{fig:RW4} (right). 
By considering  degenerate quark masses ($N_f=3$), we get the cut through the three-dimensional Columbia plot as shown in Fig.\ \ref{fig:col2} (right). All features of this plot should now be familiar.

\begin{figure}[b]
\centerline{ 
\includegraphics[height=5cm]{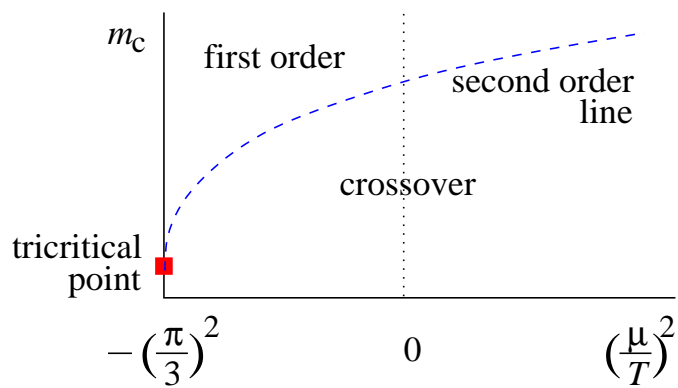}
 }
\caption{Critical quark mass separating the crossover and first-order regions as a function of $(\mu/T)^2$, for large quark masses.
}
\label{fig:mc}
\end{figure}

\begin{figure}[t]
\centerline{ 
\includegraphics[height=5.2cm]{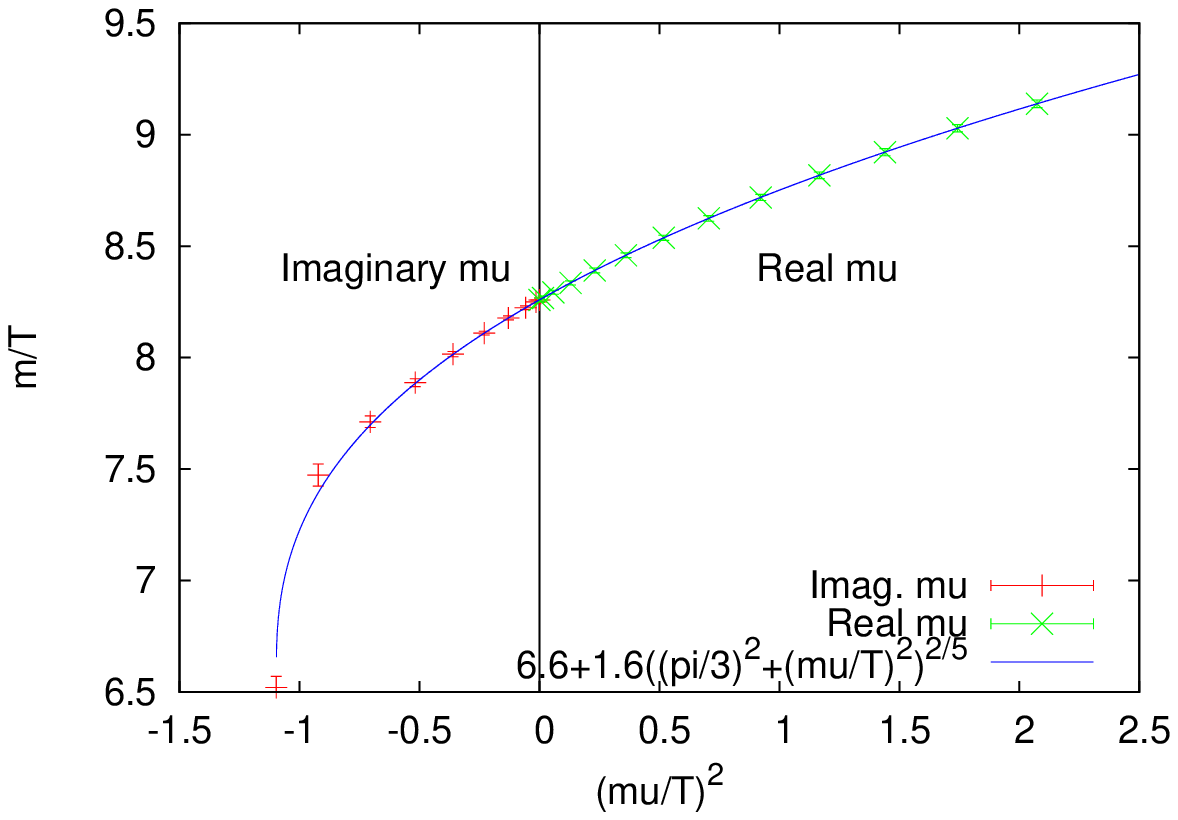} $\quad$
\includegraphics[height=5.2cm]{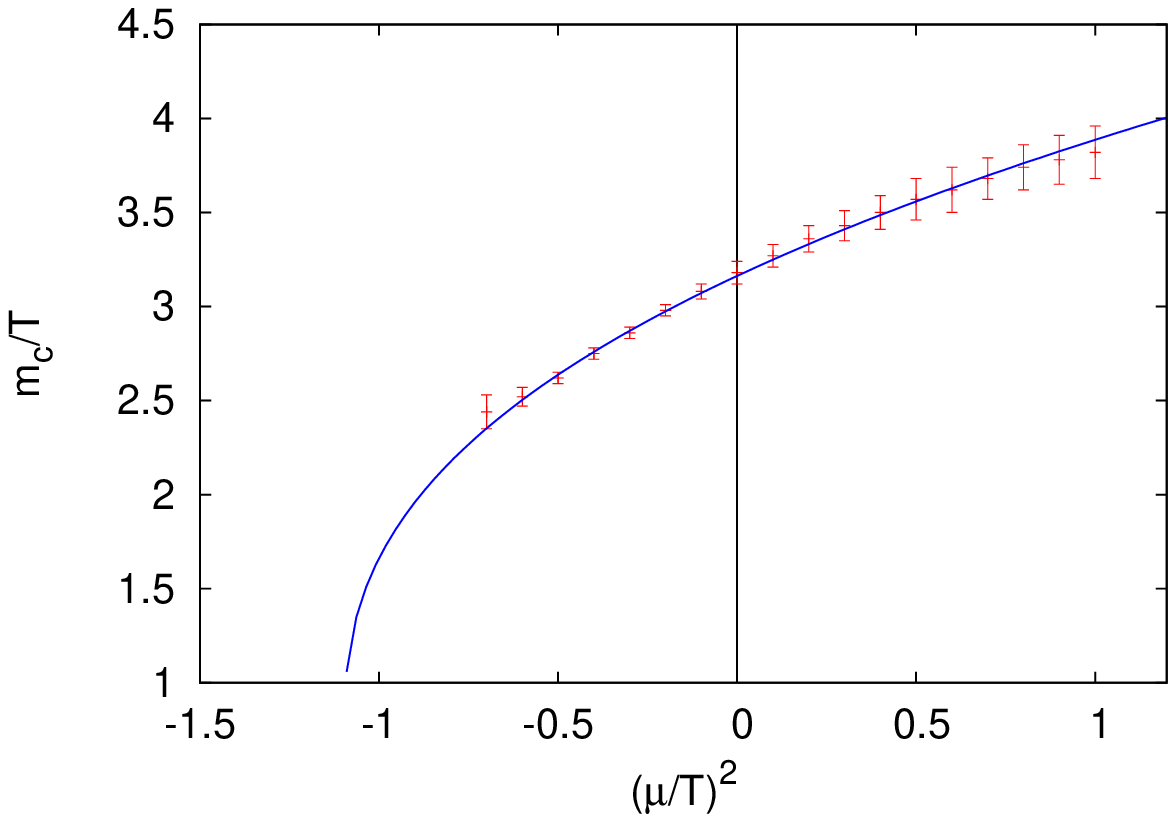}
}
\caption{Tricritical scaling \cite{deForcrand:2010he}, $m_c$ versus $(\mu/T)^2$, for heavy quarks, in the three-state Potts model \cite{Kim:2005ck} (left) and QCD in a strong coupling expansion \cite{Langelage:2009jb} (right). 
 }
\label{fig:tri}
\end{figure}

In all known cases, it appears that the first-order regions shrink as $(\mu/T)^2$ is increased. This can be made very precise for heavy quarks \cite{deForcrand:2010he,Kim:2005ck,Langelage:2009jb}. To do this, let us take the blue/dashed line on the right-hand side of Fig.\ \ref{fig:col2} (right) and rotate it. The result is shown in Fig.\ \ref{fig:mc}. The line indicates the boundary between the crossover and first-order region and is second order; we will denote it with $m_c$. Note that the line emerges from the tricritical point at $(\mu/T)^2=-(\pi/3)^2$. For increasing $\mu^2$ the first-order region shrinks, i.e.\ the critical quark mass increases. 

It turns out that the tricritical point makes its presence felt: it determines the curvature of the second-order line via tricritical scaling \cite{deForcrand:2010he}.
If we use the notation $x=(\mu/T)^2$ and $x_* = -(\pi/3)^2$, then tricritical scaling dictates that
\be
m_c(x) = m_c(x_*) + K\left( x-x_*\right)^{2/5},
\ee
where the exponent $2/5$ is fixed by universality and $K$ is a free parameter.

How well this works in practice, i.e.\ how far the scaling region extends away from $x_*$, has been tested in models where the sign problem is milder than in full QCD, namely in the three-state Potts model  \cite{Kim:2005ck}, an effective model for QCD with heavy quarks, and in QCD in a combined strong coupling and hopping parameter expansion  \cite{Langelage:2009jb}. The results are shown in Fig.\ \ref{fig:tri}.  In both models the sign problem is sufficiently mild such that simulations for real $\mu$ are possible (in the Potts model the sign problem can be eliminated completely via a reformulation \cite{Alford:2001ug} and the results for QCD actually come from semi-analytical considerations). This allows us to see that tricritical scaling works extremely well: the  data points fall on the scaling curve and the scaling region extends well into the $\mu^2>0$ domain. Hence for heavy quarks highly nontrivial predictions on the phase structure for real $\mu$ are possible from knowledge obtained purely at imaginary chemical potential, in a way
that goes substantially beyond Taylor series and analytical continuation.

To conclude, we note that tricritical scaling has been investigated mainly for heavy quarks. It is an interesting question whether this large scaling region is also present for light quarks.

\subsection{Critical endpoint}

\begin{figure}[t]
\centerline{
\includegraphics[height=4.3cm]{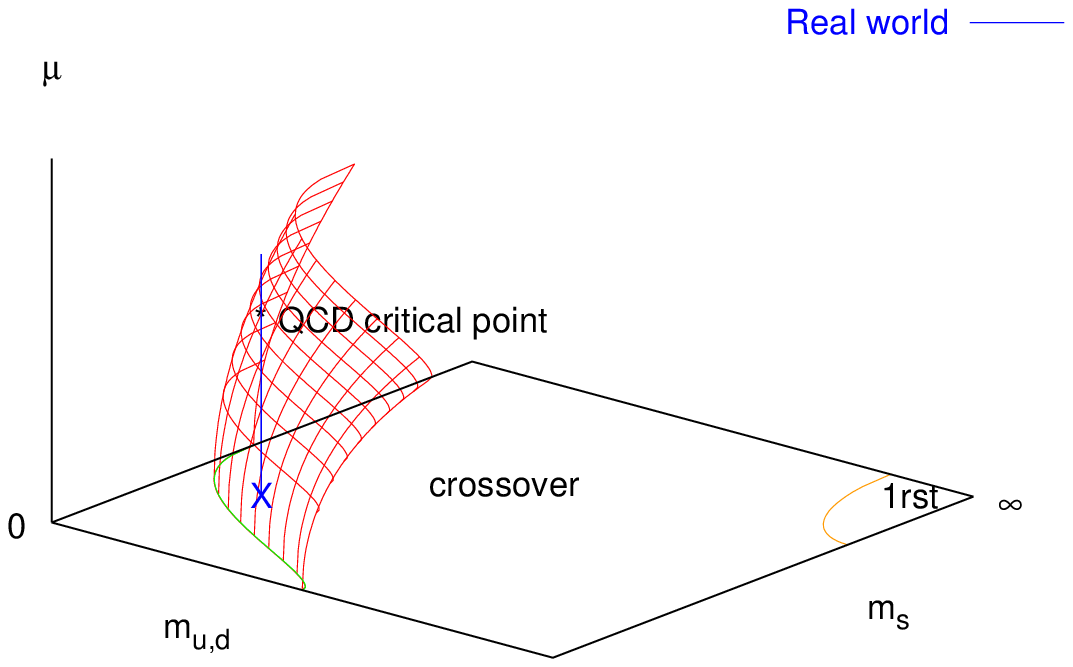} \hspace*{-1.3cm}
\includegraphics[height=4.3cm]{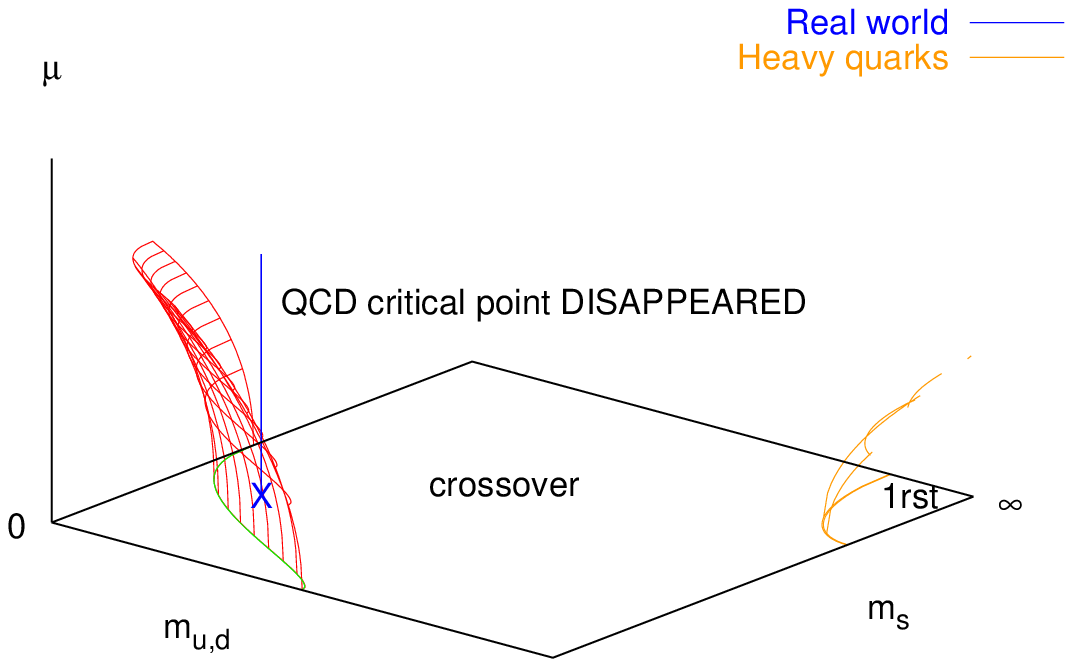} \hspace*{-1.3cm}
\includegraphics[height=4.3cm]{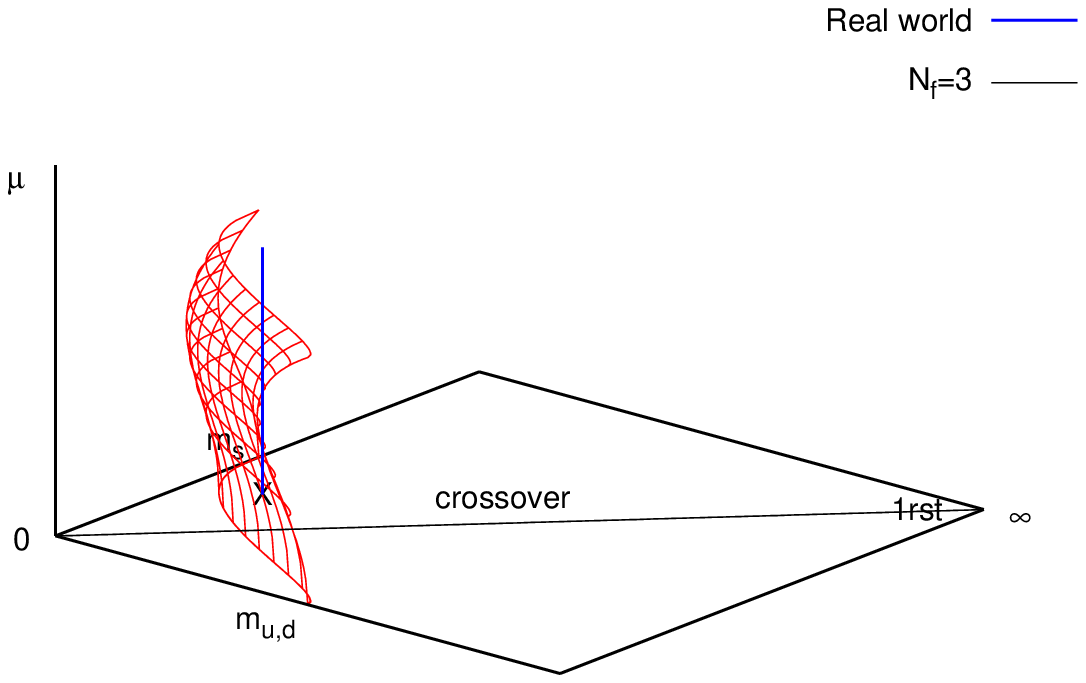}
}
\caption{Possible scenarios for the curvature of the second-order surface for light quarks and the critical endpoint for physical quark masses \cite{deForcrand:2010ys}.}
\label{fig:bend}
\end{figure}

In the three-dimensional Columbia plot the regions of first-order transitions were shown to shrink as $(\mu/T)^2$ increases. The astute reader may wonder what this implies for the critical endpoint at real chemical potential, discussed in Sec.\ \ref{sec:pb}. Here several scenarios are possible, illustrated in Fig.\ \ref{fig:bend}. Note that the position of physical quark masses is indicated with the vertical blue line. 
The standard scenario is sketched on the left: the surface bends away from the $m_q=0$ axis and the critical endpoint is located at the intersection of the (red) surface and the (blue) line.
If on the other hand the first-order region shrinks, as is the case for heavy quarks, there is no critical endpoint related to the second-order surface (centre). Finally, it is possible that the second-order surface depends in a more complicated manner on the chemical potential and quark masses, with forwards and backwards bending as $\mu$ is increased (right) \cite{Chen:2009gv}, making it substantially harder to establish its existence starting from zero or imaginary chemical potential.

This long-standing question is still not settled: it will require extensive computational resources and, ideally, approaches in which the sign problem is resolved.

\section{Complex Langevin dynamics}
\label{sec:CL}

As we have seen above, straightforward importance sampling combined with reweighting is typically  not viable, due to the overlap problem. At small $\mu/T$ it might be feasible to preserve the overlap as best as possible, on small volumes, or to use approximate methods, such as  a Taylor series expansion or analytical continuation and scaling from imaginary chemical potential. To fully attack the sign problem, however, something more radical is needed and the configuration  space should be explored in a different manner. This is what we will focus on now.

\begin{figure}[h]
\centerline{    
  \includegraphics[height=4.5cm]{plot-rho.eps}    $\qquad$
  \includegraphics[height=4.5cm]{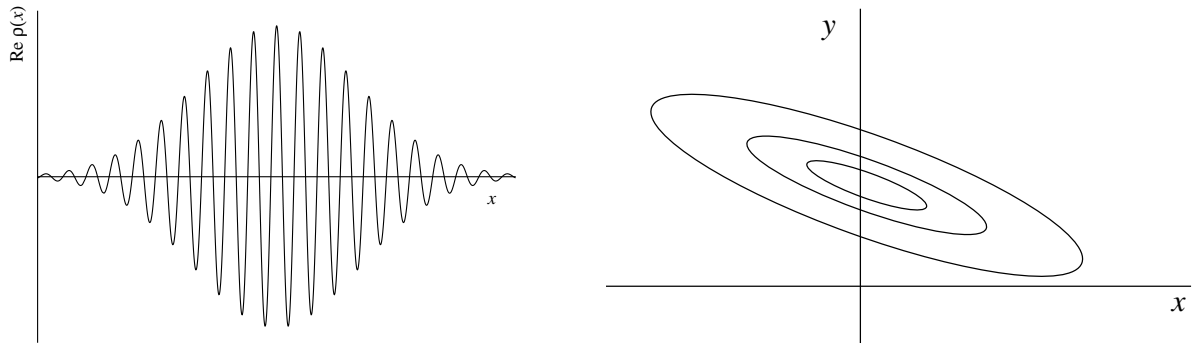}
}
\caption{What are the dominant configurations in a path integral with a complex weight? In complex Langevin dynamics, the question is answered by extending the configuration space into the complex plane.}
\label{fig:rho}
\end{figure}

The overlap problem is due to the fact that the relevant configurations differ in an essential way from those obtained at $\mu=0$ 
or using  the absolute value of the determinant. Given the excessive cancelation between configurations with `positive' and `negative' weight, one may wonder whether it is possible to give a sensible meaning to the notion of dominant configurations, e.g.\ by extending the configuration space, 
as illustrated in Fig.\ \ref{fig:rho}.
 Here we discuss the answer according to complex Langevin dynamics.

\subsection{Gaussian integrals}

We consider a simple Gaussian integral
\be
Z(a,b) = \int_{-\infty}^\infty dx\, e^{-S(x)}, \qquad\quad\quad S(x) = \half ax^2+ibx.
\label{eq:Sab}
\ee
In {\em Kindergarten}, you learn to do this integral by completing the square, i.e.\ by going into complex plane.
The lesson is therefore to analytically continue (``complexify'') the degrees of freedom, $x\to z=x+iy$, which enlarges the configuration space and gives new directions to explore. In particular, it might be possible to find a real and positive distribution $P(x,y)$, which is amenable to numerical approaches, see Fig.\ \ref{fig:rho}.

In complex Langevin dynamics, it is proposed that this distribution is constructed as the solution of a  stochastic process \cite{parisi,klauder}. To motivate this, consider again the Gaussian integral (\ref{eq:Sab}). We note that the action satisfies $S^*(b)=S(-b^*)$ and we take $a>0$ and real, such that
\be
Z(a,b) = \sqrt{\frac{2\pi}{a}} e^{-\half b^2/a}.
\ee
The corresponding phase-quenched partition function is
\be
Z_{\rm pq} =  \int_{-\infty}^\infty dx\, e^{-\half ax^2} =  Z(a,0) = \sqrt{\frac{2\pi}{a}},
\ee
and hence the average phase factor equals
\be
 \bra e^{-ibx}\ket_{\rm pq} = \frac{Z(a,b)}{Z(a,0)} = e^{-\half b^2/a}.
\ee
Since this is only a simple integral, there is no volume factor in the exponential.

The goal will be to compute expectation values, such as
\be
\bra x^2\ket = -2\frac{\partial\ln Z}{\partial a} = \frac{a-b^2}{a^2},
\ee
numerically, but without the use of importance sampling.
Let us first  take $b=0$ and use the analogy with Brownian motion \cite{Parisi:1980ys}:
a particle moving in a fluid is subject to friction ($a$) and kicks ($\eta$) and satisfies a
Langevin equation
\be
 \dot x(t) = -a x(t) + \eta(t),
\qquad\quad\quad
\bra\eta(t)\eta(t')\ket=2 \delta(t-t'). 
\ee
The kicks $\eta$ are modelled as random Gaussian noise with $\bra \eta(t)\ket=0$.
This problem is easily solved, without having to resort to numerics in this case, 
\be
 x(t) = e^{-at}x(0) + \int_0^tds\, \eta(s)e^{-a(t-s)},
\ee
and hence the correlator (taking $x(0)=0$, since the dependence on the initial condition decays exponentially in any case) equals
\be
\bra x^2(t)\ket = \int_0^tds \int_0^{t}ds' \, \bra \eta(s)\eta(s') \ket e^{-a(2t-s-s')}.
\ee
Using that $\bra\eta(s)\eta(s')\ket=2 \delta(s-s')$, we easily find
\be
 \lim_{t\to \infty} \bra x^2(t)\ket = \frac{1}{a},
\ee
which is indeed the correct answer. 

Associated with the Langevin equation is a Fokker-Planck equation for the distribution $\rho(x,t)$, defined via the relation
\be
\label{eq:FP}
\bra O(x(t)) \ket_{\eta} = \int dx\, \rho(x,t)O(x),
\ee
for a generic observable $O(x)$.
Here the noise average on the left-hand side is made explicit with the subscript $\eta$ while the average on the right-hand side is over the distribution $\rho(x,t)$. The derivation of the Fokker-Planck equation is carried out as an exercise in \ref{App:C} for the Langevin process 
\be
 \dot x(t) = K(x(t)) + \eta(t), \qquad\quad\quad K(x)=-S'(x),
\ee
where the drift $K(x)$ is derived from the action $S(x)$.
The result is
\be
\label{eq:rho}
\partial_t \rho(x,t) = \partial_x\!\left(\partial_x+S'(x)\right)\rho(x,t).
\ee
It is easy to see that the stationary solution of this Fokker-Planck  equation equals $\rho(x) \sim e^{-S(x)}$, justifying the relation (\ref{eq:FP}).
Moreover, one can show that the stationary solution is typically reached exponentially fast, see e.g.\ the comprehensive review \cite{Damgaard:1987rr}.

Let us now make the problem a bit more interesting by taking $b\neq 0$. Completing the square results in a shift in the complex plane $x\to x-ib/a$. We will now demonstrate that the same is achieved with the (complex) Langevin equation for $z=x+iy$. Writing $S(z) = S(x+iy)$, the real and imaginary parts of the Langevin equation are (see \ref{App:C} and using ``real'' noise)
\bea
\dot x =&&\hm  -\re\,\partial_z S(z) +\eta = -ax+\eta,
\\
\dot y =&&\hm  -\im\,\partial_z S(z)  =-ay-b,
\eea
 with the solution
\be
x(t) = x(0)e^{-at} + \int_0^t ds\, e^{-a(t-s)}\eta(s),
\qquad\quad
y(t) = [y(0)+b/a]e^{-at} - b/a.
\ee
The two-point correlators follow easily as
\bea
\nn
\bra x^2(t)\ket  =&&\hm x^2(0)e^{-2at} + \left(1-e^{-2at}\right)/a  \qquad\quad \to \qquad 1/a,
\\
\bra x(t)y(t)\ket =&&\hm   x(0)e^{-at}\left( [y(0)+b/a]e^{-at} - b/a\right)  \,\;\;\to  \qquad 0,
\label{eq:xy1}
\\
\nn
\bra y^2(t)\ket =&&\hm  \left([y(0)+b/a]e^{-at} - b/a\right)^2  \qquad\quad\;\;\; \to \qquad b^2/a^2.
\eea
The expressions after the arrows correspond to the limit  $t\to \infty$. 
The $n$-point functions we are interested in depend on the holomorphic combination $z=x+iy$, and we find
\be
\label{eq:xy2}
\lim_{t\to\infty}\bra (x(t)+iy(t))^2\ket = \bra x^2-y^2+2ixy\ket = 
\frac{1}{a} - \frac{b^2}{a^2} = \frac{a-b^2}{a^2},
\ee
which is as expected. We presented this in some detail to emphasise that the individual terms,  $\bra x^2\ket$, $\bra y^2\ket$ and $\bra xy\ket$, have no meaning as such: only expectation values of holomorphic observables are physically relevant.

The real and positive probability distribution associated with this process is now determined via
\be
\bra O[x(t)+iy(t)]\ket_\eta = \int dxdy\, P(x,y;t) O(x+iy),
\ee
and the Fokker-Planck equation reads
\be
\partial_t P(x,y;t) = \left[ \partial_x\left(\partial_x+\re\;\partial_z S\right)
+\partial_y \im\;\partial_z S \right] P(x,y;t).
\ee
This equation arises from a stochastic process in $x$ and $y$, but with no noise applied in the $y$ direction (see again \ref{App:C}).
However, unlike in the case of Eq.\ (\ref{eq:rho}) for $\rho(x,t)$, no generic solutions are known for this Fokker-Planck equation. In fact, even the existence of a stationary solution is not guaranteed!
This makes the justification of complex Langevin dynamics substantially harder  \cite{Aarts:2009uq,Aarts:2011ax}  than for the original real Langevin process. It relies on the equivalence of 
\be
\int dx\, \rho(x,t) O(x) = \int dxdy\, P(x,y;t)O(x+iy),
\ee
for holomorphic observables $O(x+iy)$. Refs.\  \cite{Aarts:2009uq,Aarts:2011ax} contain a detailed analysis of this problem, including consistency conditions which can be verified a posteriori.

\begin{figure}[t]
\centerline{
 \includegraphics[height=5cm]{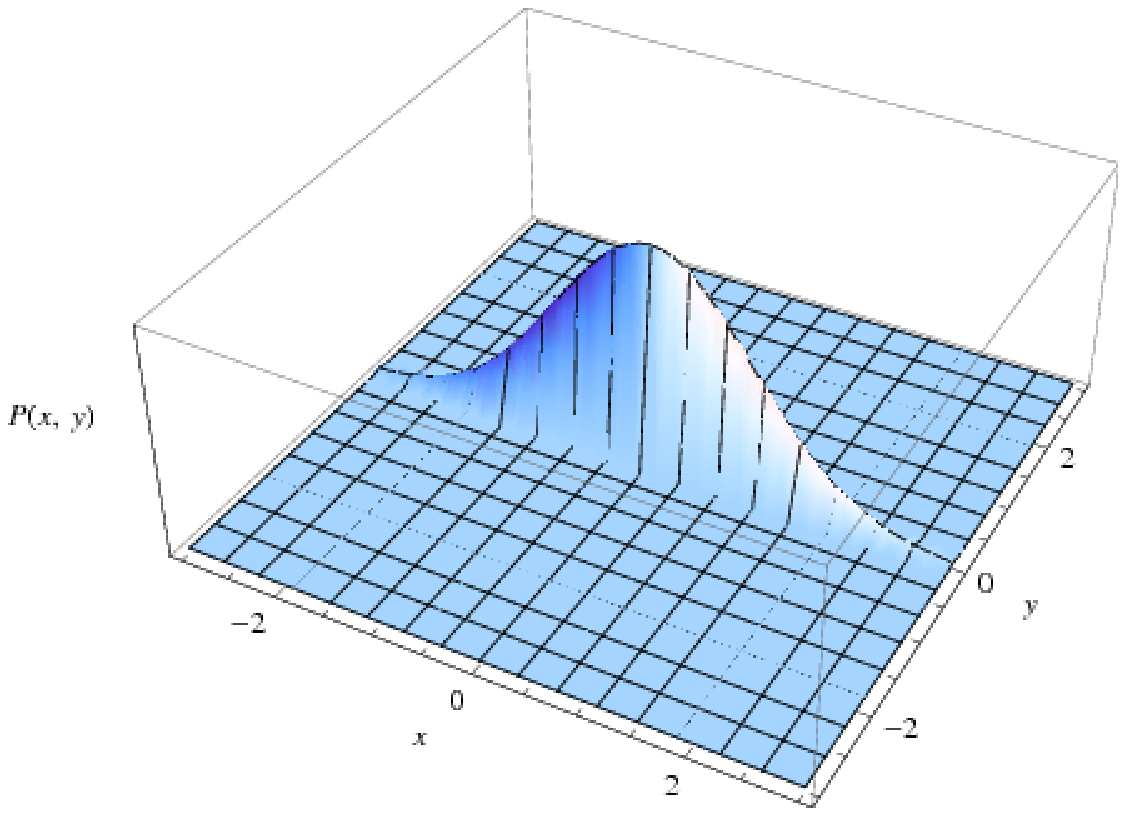}    \includegraphics[height=5cm]{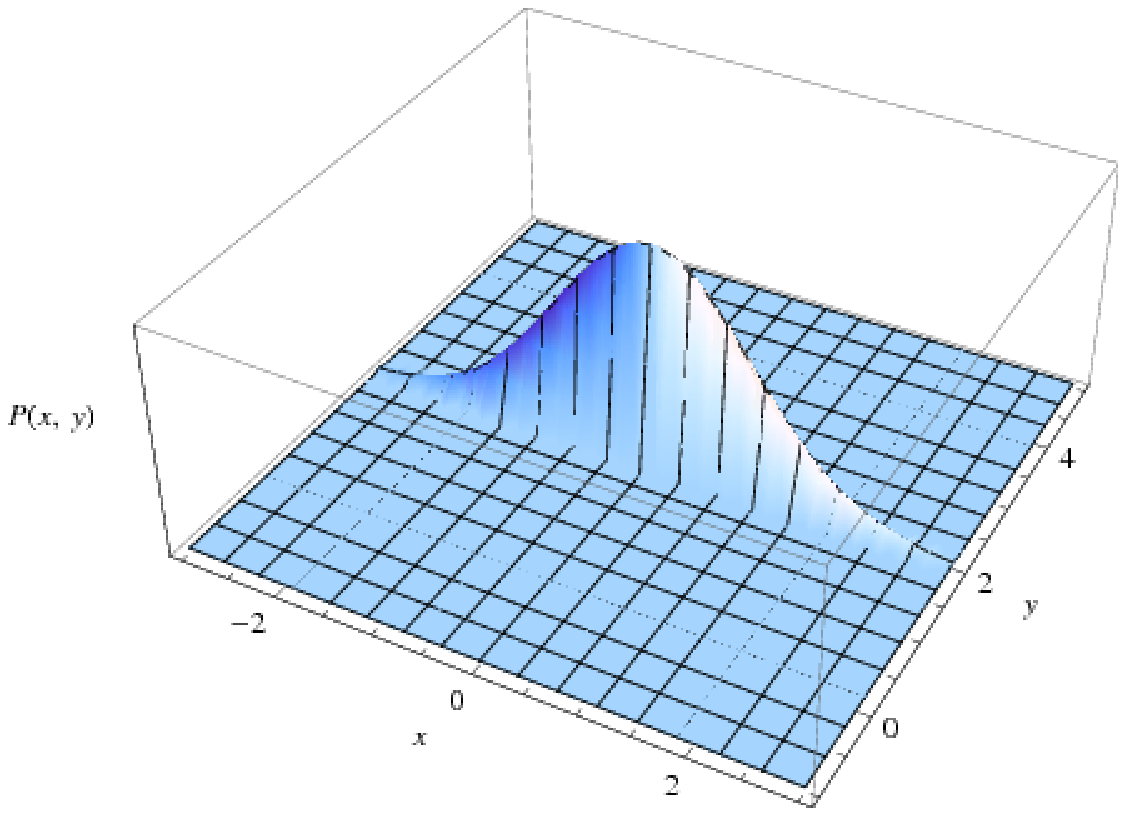}
}
\caption{Distribution $P(x,y)$ for the action $S=\half a x^2+ibx$, for $a=1$ and $b=0$ (left), $b=-2$ (right).}
\label{fig:bneq0}
\end{figure}

However, for the model considered here, the solution is easily constructed. We note that the dynamics in the $y$ direction is decoupled from $x$ and, importantly, also from the noise. Hence the distribution is simply obtained by a shift in the complex plane, $y\to -b/a$, and the distribution effectively sampled by the Langevin process equals
 \be
 P(x,y)\sim e^{-ax^2/2}\delta(y+b/a).
\ee
This is illustrated in Fig.\ \ref{fig:bneq0}. Hence the Langevin process completes the square for us.

\begin{figure}[t]
\begin{center}
  \includegraphics[height=3.6cm]{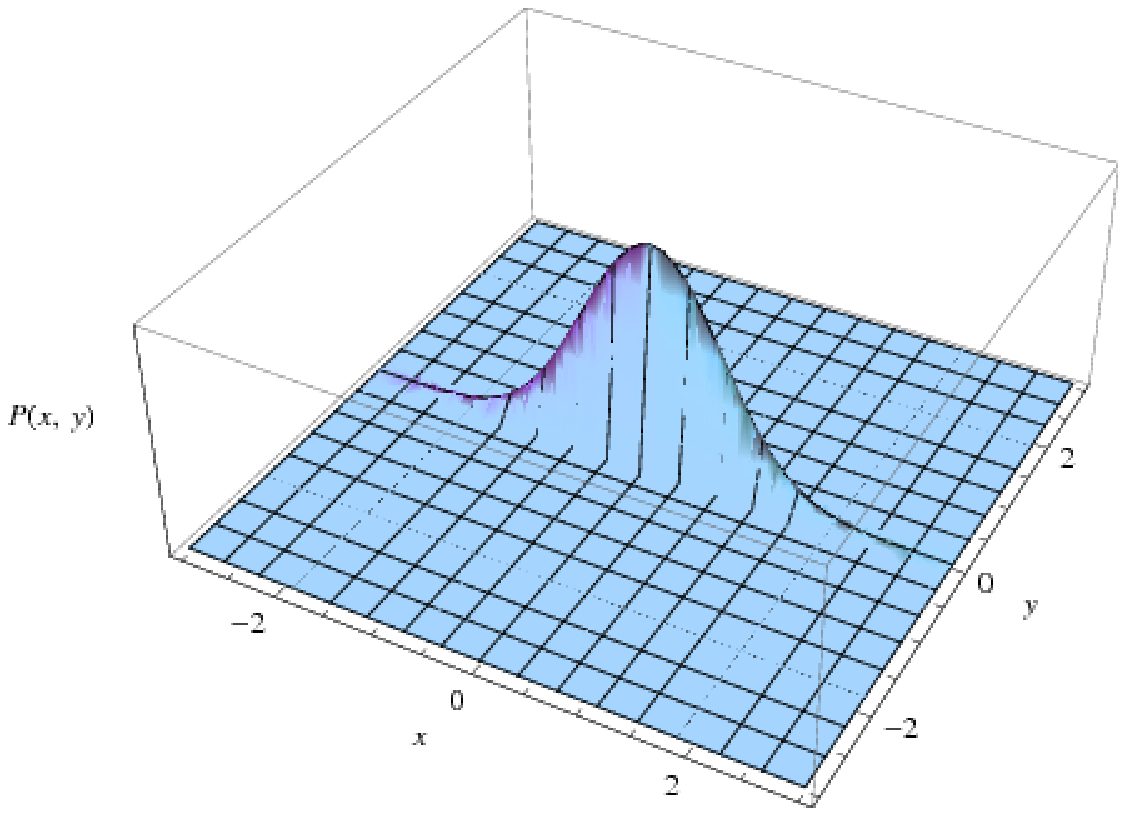}       
  \includegraphics[height=3.6cm]{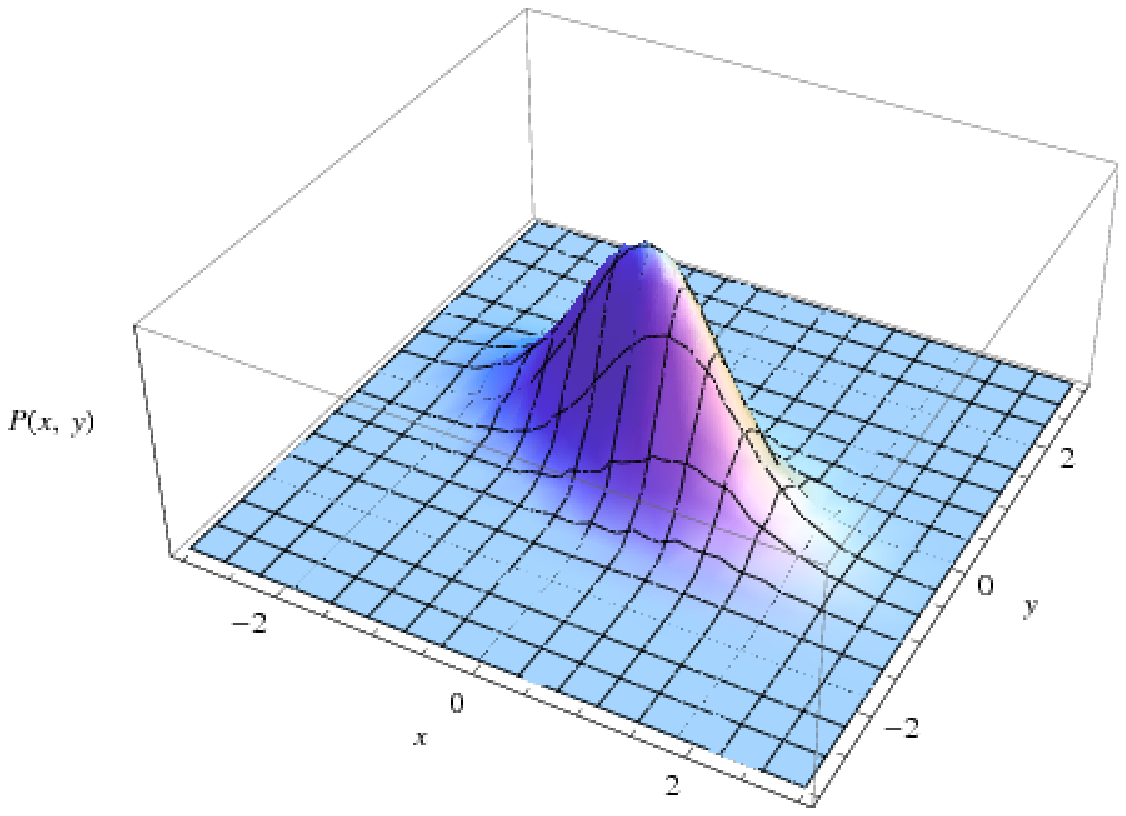}       
  \includegraphics[height=3.6cm]{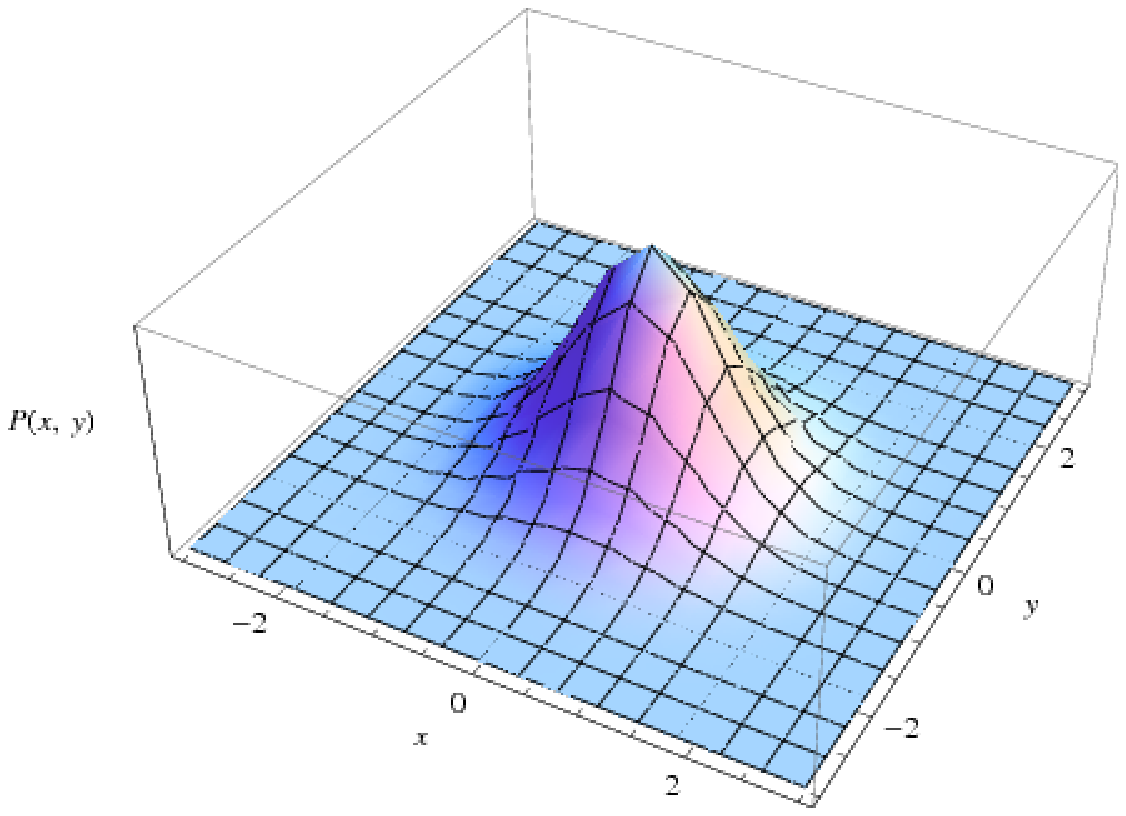}  
\end{center}
\caption{Distribution $P(x,y)$ for the action $S=\half(a+ib)x^2$ with $a=1$ and $b=0.01, 1, 10$ (from left to right).}
\label{fig:Pab}
\end{figure}

A final Gaussian example is discussed as an exercise in \ref{App:D}. In this case we consider the action
\be
S = \half(a+ib)x^2,
\ee 
and hence $x$ and $y$ are not decoupled. The resulting probability distribution $P(x,y)$ is therefore a proper two-dimensional, real and positive, distribution, as demonstrated in Fig.\ \ref{fig:Pab}. The  Langevin process finds this distribution, giving an explicit realisation of the sketch in Fig.\ \ref{fig:rho}.

\subsection{Discretisation, field theory}

Of course, most cases of interest are not analytically solvable and one has to turn to numerical solutions of the Langevin equation.
A standard (lowest-order) discretisation of the Langevin time $t=\eps n$ yields the discretised equations  \cite{Damgaard:1987rr}
\bea
\label{eq:eps}
&& x_{n+1} = x_n + \eps K^\rmR_n +\sqrt{\eps}\eta_n,  \qquad\quad K^\rmR = -\re\; \frac{\partial S}{\partial z}, 
\qquad \bra \eta_n\eta_{n'}\ket  = \delta_{nn'},
\\
&& y_{n+1} = y_n + \eps K^\rmI_n, \qquad\qquad\qquad\;\;  K^\rmI = -\im\; \frac{\partial S}{\partial z}. 
\eea
 This discretisation scheme has finite stepsize errors, which vanish linearly in $\eps$. Higher-order schemes can be used to improve the convergence to the limit of zero stepsize, see e.g.\ Ref.\ \cite{Aarts:2011zn} for an explicit example.
 Note that the stepsize can be chosen adaptively if necessary \cite{arXiv:0912.0617}. 
 
In field theory, the approach outlined here is known, for real actions, as stochastic quantisation  \cite{Parisi:1980ys} and, for complex actions or drifts,
 as complex Langevin dynamics  \cite{parisi,klauder}.
 We consider the euclidean path integral  
\be
 Z = \int D\phi\, e^{-S}.
 \ee
Now Langevin dynamics takes place in the `fifth' time direction, as
\be
 \frac{\partial\phi(x,t)}{\partial t} = -\frac{\delta
S[\phi]}{\delta
\phi(x, t)} + \eta(x, t),
\ee
with the noise satisfying
\be
 \bra \eta(x, t)\ket = 0
 \qquad\quad
 \bra \eta(x, t)\eta(x', t')\ket = 2\delta(x-x')
\delta( t- t').
\ee
As in the cases above, one computes expectation values $\bra\phi(x, t)\phi(x', t)\ket$, etc, and studies the convergence as the Langevin time 
 $ t\to\infty$. Both the discretisation and the complexification are as above. Gauge theories will be discussed in the next section.

\subsection{Applicability to theories with a sign problem}

Langevin dynamics for real actions -- stochastic quantisation -- was discussed extensively in the 1980s and equivalence with path integral quantisation has been demonstrated \cite{Damgaard:1987rr}. For complex actions on the other hand, a formal proof was notably absent. This situation has improved considerably in recent years and the theoretical foundation has now been formulated \cite{Aarts:2009uq}. Moreover, practical criteria for correctness can be written down and these can be assessed in numerical studies \cite{Aarts:2011ax}.
 Failure of the approach, first observed in the 1980s \cite{Ambjorn:1985iw,Ambjorn:1986fz}, can be explained within the theoretical framework, albeit only a posteriori, see e.g.\ Ref.\ \cite{arXiv:1005.3468} for a discussion of success and failure in the three-dimensional XY model with a complex action.

 A crucial role in the justification is played by the distribution $P(x,y)$. Provided that the action is holomorphic and  the distribution $P(x,y)$ is sufficiently localised, i.e.
\be
P(x,y)=0 \;\; \mbox{for} \;\; |y|>y_{\rm max} \qquad \mbox{[or} \;\; P(x,y)\to 0 \;\; \mbox{fast enough],}
\ee
correct results are obtained, modulo some technical requirements \cite{Aarts:2009uq,Aarts:2011ax}.
The proof relies  on the Cauchy-Riemann equations, and hence holomorphicity of the drift and observables, and the possibility to perform partial integral in the imaginary direction without picking up boundary conditions.
In the case of  simple models with a holomorphic action, this has led to a complete, both numerical and analytical, understanding \cite{Aarts:2013uza}.

An open question concerns the situation with meromorphic drifts, i.e.\ drifts with poles, which arise for instance from the inclusion of a log det
in the effective action, schematically
\be
Z = \int dx\, e^{-S} \det M = \int dx\,e^{-S_{\rm eff}}, \qquad\quad S_{\rm eff}=S-\log \det M,
\ee
with a drift
\be
K = -\partial_z S_{\rm eff} = -\partial_z S +\Tr M^{-1}\partial_z M.
\ee
In this case the assumed holomorphicity of the (effective) action is not present and the formal derivation has to be reconsidered.
In practice, it has been found that problems {\em may} appear but not necessarily so \cite{Mollgaard:2013qra}.
This is still very much a topic of ongoing studies 
\cite{Aarts:2014bwa,Mollgaard:2014mga,Splittorff:2014zca,Nishimura:2015pba}.

Aside from this important issue, the most exciting findings are that it has been shown that the method can handle severe sign and Silver Blaze problems, e.g.\ in the four-dimensional Bose gas at nonzero chemical potential \cite{Aarts:2008wh}. The applicability of complex Langevin dynamics to the SU(3) spin model, to which it was first applied in 1985 \cite{KW}, is now also understood \cite{Aarts:2011zn}, as has the essential difference between abelian and nonabelian spin models \cite{Aarts:2012ft}.
Most importantly, in the context of the QCD, there has been essential progress in the treatment of nonabelian gauge theories, to which we turn now.

\section{Complex Langevin dynamics for gauge theories}
\label{sec:CL2}

The recent interest in complex Langevin dynamics for QCD at nonzero density arises from successful applications to SU(3) gauge theory, first in the presence of heavy (static) quarks and then also in the presence of dynamical quarks. This is still very much a topic in development, so in these lectures I want to focus on the algorithmic  advance of gauge cooling \cite{Seiler:2012wz}, which, in combination with the improved analytic understanding mentioned above, has led to some remarkable progress.

In SU($N$) gauge theories, the complexification works as follows \cite{Berges:2006xc,Aarts:2008rr}. Originally the gauge links $U_{x\nu}$ are elements of SU($N$), i.e., they are unitary with determinant 1.
After discretisation of the Langevin time and using a lowest-order scheme in $\eps$,  a (complex) Langevin update takes the form \cite{Batrouni:1985jn}, \be
 U_{x\nu}(n+1)  = R_{x\nu}(n)\, U_{x\nu}(n),
\qquad\quad\quad
R_{x\nu} = \exp \left[ i\lambda_a\left( \eps K_{x\nu a} +\sqrt \eps \eta_{x\nu a}\right)  \right].
\ee
Here $\lambda_a$ are the Gell-Mann matrices  and a sum over the indices $a=1,\ldots N^2-1$ is assumed.
$K_{x\nu a}$ is the drift,
  \be
  K_{x\nu a} =  -D_{x\nu a} (S_{\rm YM}+S_{\rm F}),
\qquad\quad\quad S_{\rm F} = - \ln\det M,
\ee
where the action includes the logarithm of the fermion determinant. Differentiation is defined as
\be
D_{x\nu a} f(U) = \frac{\partial}{\partial\alpha} f\left(e^{i\alpha \lambda_a}U_{x\nu}\right)\Big|_{\alpha=0},
\ee
and the noise is normalised as usual,
\be
\left\bra \eta_{x\nu a}(n)\eta_{x'\nu' a'}(n')\right\ket = 2\delta_{xx'}\delta_{\nu\nu'}\delta_{aa'}\delta_{nn'}.
\ee
Below we will suppress indices, when appropriate.
Note that the combination of the drift and the noise in the exponential appears just as in Eq.\ (\ref{eq:eps}) for one degree of freedom. Since the Gell-Mann matrices are traceless, the determinant of $R$ and hence of $U$ remain 1 for any choice of $K$ and $\eta$.
Moreover, if the action and therefore the drift $K$ are real, $R$ and $U$ will remain unitary, using this update.

Let us now consider the case that the action (or the fermion determinant) is complex. In that case $K^\dagger \neq K$ and $U$ will no longer be unitary. Instead,  $U$ will take values in the special linear group, i.e.\ the complexification in this case is from SU($N$) to SL($N,\mathbb{C}$). One remark is that now $U^\dagger$ and $U^{-1}$ are no longer identical. Since complex Langevin dynamics provides the analytical continuation of the original theory, it is essential that links are written as $U$ and $U^{-1}$ in the action, such that $S(U)$ is a holomorphic function of $U$ in principle (ignoring possible issues due to the fermion determinant here). For instance, the original statement of unitarity, $UU^\dagger=\id$, is now replaced with $UU^{-1}=\id$, which still holds of course. Similarly, physical observables should be written as functions of $U$ and $U^{-1}$, such that they are holomorphic. This is similar to the discussion for the Gaussian models above, see e.g.\ Eqs.\ (\ref{eq:xy1}, \ref{eq:xy2}).

\begin{figure}[t]
  \centerline{   
  \includegraphics[height=5.7cm]{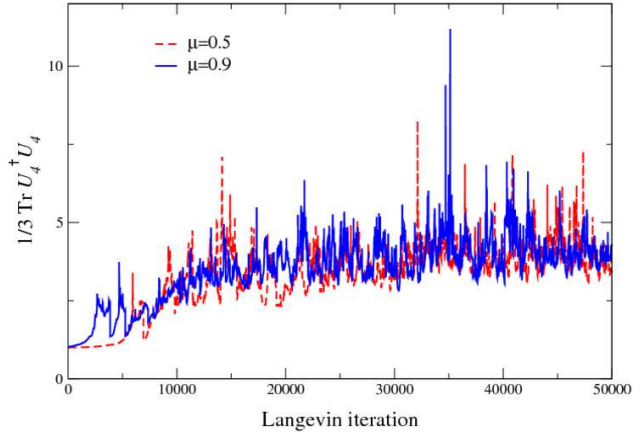}
  }
  \caption{Deviation from SU($3$): Langevin time evolution of the unitarity norm $\Tr U_4^\dagger U_4/3\geq 1$ in heavy dense QCD on a $4^4$ lattice with $\beta=5.6$, $\kappa=0.12$, $N_f=3$ \cite{Aarts:2008rr}.
}
\label{fig:d}
\end{figure}

On the other hand, nonholomorphic combinations can be used to gain insight in the complex Langevin process. In the Gaussian models, the width of the distribution $P(x,y)$ is given by $\bra y^2\ket$, or $\bra y^k\ket$ in general. Similarly, the deviation from SU($N$) can be studied using so-called unitarity norms \cite{Aarts:2008rr,Seiler:2012wz,Aarts:2013uxa}, such as
\be
\frac{1}{N}\Tr \left(UU^\dagger -\id\right) \geq 0, 
\qquad\quad\quad
\frac{1}{N}\Tr \left(UU^\dagger -\id\right)^2 \geq 0,
\ee
etc,
where the second inequality is obvious and the first inequality follows from the polar decomposition of $U\in$ SL($N,\mathbb{C}$): $U = VP$, with $V\in$ SU($N$) and $P$ a positive semidefinite hermitian matrix with $\det P = 1$ \cite{Aarts:2008rr}. 
Indeed, during a complex Langevin simulation these norms become nonzero, as demonstrated in Fig.\ \ref{fig:d} for QCD in the presence of static quarks for two values of the chemical potential $\mu$.

Intuition says that during a simulation the evolution should be controlled in the following way: configurations should stay close to the SU($N$) submanifold 
\begin{itemize}
\item when the chemical potential $\mu$ is small;
\item when small nonunitary initial conditions are introduced;
\item in the presence of roundoff errors.
\end{itemize}
In practice however, it turns out that this is not the case and the unitary submanifold is unstable. This observation has been made one way or the other several times in the past, but the relation between this and the breakdown of the approach -- convergence to incorrect results -- is fairly recent and follows from the combination of theoretical and numerical ideas 
\cite{Aarts:2009uq,Aarts:2011ax,Seiler:2012wz}.
Given what we learnt in previous sections, a simple way to test whether this instability arises is to study analyticity (or lack thereof) of observables around $\mu^2\sim 0$.

\subsection{Gauge cooling}

The instability of the SU($N$) submanifold is related to gauge freedom. Consider a link at site $k$, which transforms as
\be
U_k\to \Omega_k U_k\Omega_{k+1}^{-1},
\qquad\quad\quad
\Omega_k = e^{i\om_a^k\lambda_a},
\ee
with $\om_a^k$ the gauge parameters. Note that in SU($N$), $\om_a^k\in \mathbb{R}$, while in SL($N,\mathbb{C}$), $\om_a^k\in \mathbb{C}$. While unitary gauge transformations preserve the unitarity norms, SL($N,\mathbb{C}$) transformations with $\om_a^k$ nonreal do not. In fact, those transformations can make the unitarity norms increase out of bounds, leading to broad undesirable distributions. 

\begin{figure}[t]
 \begin{center}
 \includegraphics[height=5cm]{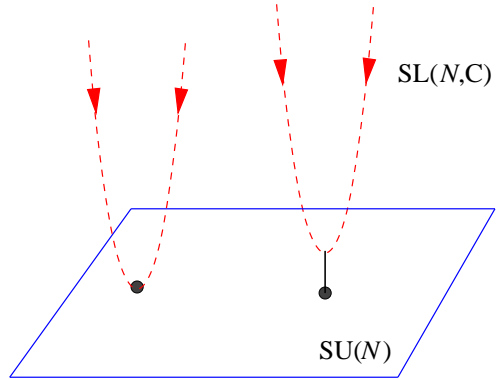} 
 \end{center}
\caption{Gauge cooling of links in SL($N, \mathbb{C}$) reduces the distance from SU($N$). The left orbit is equivalent to a SU($N$) configuration, while the one on the right is not \cite{Aarts:2013uxa}.
}
\label{fig:sl}
 \end{figure}

Having made this observation, one can use it in a constructive manner. It is possible to devise gauge transformations that {\em reduce} the unitarity norms and hence control the Langevin evolution. This goes under the name {\em gauge cooling}  \cite{Seiler:2012wz}. We hence consider
\be
U_k \to \Omega_k U_k \Omega_{k+1}^{-1}, 
\qquad\quad\quad 
\Omega_k = e^{-\alpha f_a^k\lambda_a},
\qquad\quad \alpha>0,
\ee
or, similarly, a cooling update at site $k$, 
\be
U_k\to \Omega_k U_k, \qquad\quad\quad  U_{k-1}\to U_{k-1}\Omega_k^{-1}.
\ee
Let us consider the effect of this on the unitarity norm
\be
\dd = \sum_k \frac{1}{N}\Tr \left(U_kU_k^\dagger - \id\right).
\ee
After one update and linearising in $\alpha$, we find
\be
\dd '-\dd =  -\frac{\alpha}{N} (f_a^k)^2 +{\cal O}(\alpha^2) \leq 0,
\ee
i.e. the distance from SU($N$) has indeed been reduced, see Fig.\ \ref{fig:sl}.

Up to now, we have not defined $f_a^k$. One possibility is to choose it as the gradient of the unitarity norm itself, i.e. 
\be
f_a^k = 2\Tr\lambda_a\left( U_kU_k^\dagger -U_{k-1}^\dagger U_{k-1}\right).
\ee
 Hence when $U\in$ SU($N$), we find that $f_a^k=0$ and cooling has no effect, as it should be. 

We will now demonstrate this in a simple one-link example \cite{Aarts:2013uxa}, with the action and gauge freedom,
\be
 S = \frac{1}{N}\Tr U,
\qquad\quad\quad
U\to \Omega U\Omega^{-1}.
\ee
The distance $\dd$ and gauge cooling function $f_a$ are 
\be
\dd = \frac{1}{N}\Tr \left(UU^\dagger-\id\right),
\qquad\quad\quad
f_a = 2\Tr\lambda_a\left( UU^\dagger -U^\dagger U\right).
\ee
In this simple model, the trace of $U$ is invariant under cooling and hence $c=\Tr U/N$ and $c^*=\Tr U^\dagger/N$  are preserved.
After one cooling update, we find that
\be
\dd' - \dd = -\frac{\alpha}{N} f_a^2 = -\frac{16\alpha}{N} \Tr UU^\dagger[U,U^\dagger].
\ee
Specialising now to SU(2) and SL(2,$\mathbb{C}$), while taking the continuous cooling-time limit, $\dd' - \dd\to \dot\dd$, we find the cooling equation
\be
\label{eq:cooling}
\dot \dd =  -8 \alpha \left( \dd^2+2\left(1-|c|^2\right)\dd+c^2+c^{*2}-2|c|^2\right),
\ee
expressed in terms of the invariants $c$ and $c^*$. We can now consider two cases:
\begin{itemize}
\item $c=c^*$: $U$ is gauge equivalent to an SU(2) matrix. Eq.\ (\ref{eq:cooling}) simplifies to 
\be
\dot\dd = 8\alpha(\dd+2-2c^2)\dd,
\ee
with the asymptotic solution
\be
\dd(t) \sim e^{-16\alpha(1-c^2)t} \to 0,
\ee
i.e.\ the distance vanishes, as expected.
\item $c\neq c^*$: $U$ is not gauge equivalent to an SU(2) matrix. Hence there is a minimal distance from the SU(2) submanifold, given by the fixed point of  Eq.\ (\ref{eq:cooling}),
\be
\dd(t)\to \dd_0 =  |c|^2-1+\sqrt{1-c^2-c^{*2}+|c|^4}>0,
\ee
which is again reached exponentially fast.
\end{itemize}
This demonstrates the idea behind gauge cooling in a simple, analytically solvable example. It should be noted that for many links the approach to the minimal distance is no longer exponential but appears to follow a power law \cite{Aarts:2013uxa}.
Finally, the parameter $\alpha$ can be chosen adaptively, to optimise the numerical implementation \cite{Aarts:2013uxa,Aarts:2014kja}.

\subsection{Complex Langevin dynamics with gauge cooling for QCD}

In QCD the unitary submanifold is unstable. Even in SU(3) gauge theory, without a complex action, links will not remain unitary when Langevin updates are employed, due to roundoff errors. These are of course also present in other algorithms, but for simulations of theories with a real action they can be easily controlled, namely by occasionally re-unitarising the links, i.e.\ projecting them back into SU(3). However, this option is not available when the action is complex and links {\em should} be outside of SU(3). Hence, it becomes necessary to use an alternative manner to control the exploration and this is provided by gauge cooling. In practice Langevin updates and cooling updates are alternated, with considerable freedom in the number of cooling steps, and choosing both the gauge cooling parameter and the Langevin stepsize adaptively. 

This approach \cite{Seiler:2012wz} was first applied to heavy dense QCD, i.e.\ SU(3) gauge theory in the presence of heavy quarks. Since the (anti)quarks are static, they are represented by (conjugate) Polyakov loops, and the fermion determinant for a single flavour takes on a simple form \cite{Aarts:2008rr},
\be
\label{eq:HD}
\det M = \prod_\xv \det\left(1+h e^{\mu/T} {\cal P}(\xv) \right)^2 \left(1+h e^{-\mu/T} {\cal P}^{-1}(\xv)\right)^2,
\ee
where the remaining determinant is in colour space only, ${\cal P}$ is the untraced Polyakov loop,
\be
{\cal P}(\xv) =  \prod_{\tau=0}^{N_\tau-1} U_4(\tau,\xv),
\ee
and $h=(2\kappa)^{N_\tau}$ is related to the quark mass via $m_q=-\ln(2\kappa)$ at leading order in the hopping expansion considered here.
The determination of the phase diagram using this approach is in progress and a recent status  report can be found in Ref.\ \cite{Aarts:2015yba}.
It should be noted that gauge cooling stops being effective when the gauge coupling $\beta$ is chosen too small  \cite{Seiler:2012wz}, i.e.\ on a coarse lattice. Since ultimately one has to reduce the lattice spacing (increase the gauge coupling $\beta$), this is no problem in principle, but it rules out tests on small lattices. It also stimulates the search for alternatives to gauge cooling.

The first application to full QCD, i.e.\ with dynamical quarks, can be found in Ref.\ \cite{Sexty:2013ica}. This constitutes a major step forward. In Ref.\ \cite{Aarts:2014bwa}
  results in full QCD were subsequently compared with those obtained in QCD using a hopping parameter expansion to all orders, and  in Ref.\ \cite{Fodor:2015doa} with reweighting.
  Finally, complex Langevin dynamics has also been applied to SU(3) gauge theory in the presence of a nonzero $\theta$-term \cite{Bongiovanni:2014rna}.

As stated above, this topic is in continuous development and hence it is premature to discuss physically relevant results in detail at this stage.
Two outstanding problems concern the treatment of poles in the drift, present due to the logarithm of the fermion determinant in the effective action, and the breakdown of gauge cooling at small $\beta$.

\section{Other approaches}
\label{sec:other}

There are more approaches under development  to tackle the sign problem than I was able to cover in the lectures.
This final section contains a partial list of other methods that are currently being studied and does not aspire to be complete. Its main message should be the variety of ideas and proposals that are available and hence the richness of the subject.
Some of these proposals, and some others, are reviewed a bit more extensively in Refs.\ \cite{Aarts:2013naa,Aarts:2015kea}.

\subsection{Changing the order of integration, strong coupling}

As we saw throughout these lectures, it is the complex fermion determinant at nonzero chemical potential that leads to the sign problem. 
Hence it makes sense not to integrate out the fermions first but instead perform the integral over the gauge links. 
Since the gauge sector is an interacting theory by itself, this cannot be done exactly. One possible starting point is to consider the strong-coupling limit, where the gauge coupling $\beta=2N_c/g^2$, the coefficient of the plaquette in the (Yang-Mills) action, is taken to zero. The gluonic path integral now  factorises into the product of one-link integrals, which can be done analytically. The resulting partition function has a very physical representation in terms of worldlines of mesons and baryons (Monomer-Dimer-Polymer (MDP) system) \cite{Karsch:1988zx}, which can be studied \cite{deForcrand:2009dh,Unger:2011it} using efficient worm-type algorithms \cite{Prokof'ev:2001zz}.
In order to go beyond the strong-coupling limit, one may include the first ${\cal O}(\beta)$ corrections \cite{deForcrand:2014tha}, or include the plaquettes by integrating them in steps, via the introduction of auxiliary fields \cite{Fabricius:1984wp,Vairinhos:2010ha,Vairinhos:2014uxa,Brandt:2014rca}. 

Another  possibility is to combine the strong coupling expansion with the hopping expansion for quarks (an expansion in the inverse mass), to construct effective models amenable to numerical simulations \cite{Fromm:2011qi}.
The determinant at leading order in the hopping expansion was already presented in Eq.\ (\ref{eq:HD}). Going to higher order cures a number of deficits of the static limit and gives access to the onset to cold dense matter for heavy quarks \cite{Fromm:2012eb}.

\subsection{Dual formulations}

A related approach uses a strong-coupling expansion to all orders. This method is not easily applicable to nonabelian gauge theories, but has been very successful in abelian theories and spin models. For illustration, consider the three-dimensional  SU(3) spin model, an effective model for QCD at nonzero temperature and density  \cite{KW}.  The action is written as
\be
S=S_B+S_F, 
\ee
with
\be
\label{eq:SU3}
 S_B = -\beta\sum_{\bra x y\ket} \left[ P_x P_y^* + P_x^* P_y\right],
 \qquad\quad\quad
 S_F = -h\sum_x\left[ e^\mu P_x+e^{-\mu} P_x^*\right].
 \ee
 The degrees of freedom are effective Polyakov loops, $P_x=\Tr U_x$, $P_x^*=\Tr U_x^\dagger$, where the $U_x$'s are SU(3) matrices, living on a three-dimensional lattice. Static (anti)quarks are represented by (conjugate) Polyakov loops, weighted with the chemical potential to introduce an imbalance. Note that $S^*(\mu)=S(-\mu^*)$. Expanding the Boltzmann weight to all orders, as in a classical high-temperature expansion, yields a representation of the partition function containing terms of  the following form \cite{Gattringer:2011gq}

 \be
I(n_x, \bar n_x) = \int_{\rm SU(3)} dU_x\, \left(\Tr U_x\right)^{n_x} \left(\Tr U_x^\dagger\right)^{\bar n_x}.
\ee
Crucially, it is now possible to perform these single-site SU(3) integrals, yielding again a monomer-dimer system with constraints, since $I(n_x, \bar n_x)$ is only nonzero provided $\left(n_x-\bar n_x\right)$ mod 3 $=0$. It turns out that all the nonzero weights that appear in this representation are real and positive, even when $\mu\neq 0$. These, and similar, models can then be solved with importance sampling or a worm algorithm, see 
Ref.\  \cite{Gattringer:2014nxa} for a review.

\subsection{Density of states, histograms}

In the approach known as density of states,  factorisation, histogram method or Wang-Landau
\cite{Gocksch:1988iz,Ambjorn:2002pz,Fodor:2007vv,Ejiri:2007ga,Anagnostopoulos:2010ux,Ejiri:2013lia,Saito:2013vja,Langfeld:2012ah},
 the idea is to evaluate the path integral in two stages. At the first stage one evaluates a constrained integral with one degree of freedom fixed. At the second stage the remaining integral over the resulting probability distribution -- the density of states -- constructed in the first step, is performed. The density of the states can be obtained by constructing histograms during the constrained simulation.  For example, if $P$ denotes  the degree of freedom that is kept fixed (e.g.\ the plaquette, action density or  Polyakov loop), then the (unnormalised) density of states is given by
\be
\label{eq:w}
w(P) =  \int DU\, \delta(P-P') \, e^{-S_{\rm YM}} \det M,
\ee
where $P'$ is the value of $P$ taken during the simulation. The expectation value of $P$ is then determined by the simple integrals
\be
\bra P\ket = \frac{1}{Z} \int dP\, w(P) P,
\quad\quad\quad\quad
Z= \int dP\, w(P).
\ee
The main issues in this approach are 
\begin{enumerate}
\item the contrained integral should have a positive weight, so that it can be determined unambiguously;
\item  the weight $w(P)$ should be computable to very high {\em relative} precision;
\item the remaining integral should be doable, which may be nontrivial due to the sign problem.
\end{enumerate}
Recently promising results have been obtained with  improvements  \cite{Langfeld:2012ah,Langfeld:2013xbf,Langfeld:2014nta,Gattringer:2015lra} of the Wang-Landau algorithm \cite{WL}.

\subsection{Lefschetz thimbles}

In complex Langevin dynamics, a complexified configuration space is explored, relying on holomorphicity of the theory under consideration. This arises naturally in the simpler integrals considered earlier, which are often evaluated using steepest-descent or stationary-phase approximations, with saddle points in the complex plane. This can be made mathematically very precise using integration along so-called Lefschetz thimbles 
\cite{Witten:2010cx}. The numerical implementation of this idea for QCD and other field theories was proposed in Ref.\ \cite{Cristoforetti:2012su}.

In this  approach the integration path is deformed such that it passes through the fixed (or critical) points of the complex action.  The integration contour follows paths of steepest descent, along which the imaginary part of the action is constant; these are the (stable)  thimbles ${\cal J}$. 
For one degree of freedom and one saddle point, this amounts to writing
\bea
Z =&&\hm \int dx\, e^{-S(x)} = e^{-i\im\, S_{\cal J}} \int_{\cal J} dz\, e^{-\re\, S(z)}
\nn\\
=&&\hm  e^{-i\im\, S_{\cal J}} \int ds\, J(s) e^{-\re\, S(z(s))},
\quad\quad\quad J(s) = x'(s)+iy'(s).
\eea
Since the imaginary part of the action is constant along the thimble, it can be taken out of the integral.
Two sign problems remain in this formulation: in the second line we have parametrised the thimble in terms of $z(s)=x(s)+iy(s)$, resulting in a complex jacobian $J(s)$ in general. This yields a residual sign problem, which may however be milder than the original one  \cite{Cristoforetti:2012su}. The second sign problem arises if more than one critical point contributes and the associated thimbles differ in their phases. The partition function is then a sum over thimbles and
\be
Z = \sum_k  m_k e^{-i\im\, S_{{\cal J}_k}} \int_{{\cal J}_k} dz\, e^{-\re \,S(z)},
\ee
with $m_k$ the so-called intersection number. How to treat these global phases numerically is not clear yet, but based on universality
 it has been conjectured that a single saddle point (e.g.\ the perturbative one) suffices \cite{Cristoforetti:2012su}.
This is often not the case in simpler models, but it should be noted that these models will typically lack universality.
Lefschetz thimbles for lattice models are currently actively being studied, from a number of angles:
 residual sign problem and numerical algorithms \cite{Cristoforetti:2013wha,Mukherjee:2013aga,Fujii:2013sra,Cristoforetti:2014gsa,Alexandru:2015xva};  thimble structure and the global sign problem in simple models
\cite{Kanazawa:2014qma,Tanizaki:2015rda,Fujii:2015bua}; and
a comparison with complex Langevin dynamics \cite{Aarts:2013fpa,Hayata:2015lzj}, including the 
 role of zeroes of the determinant \cite{Aarts:2014nxa}.

\section{Conclusion}
\label{sec:conc}

In these lectures a basic introduction to lattice QCD at nonzero chemical potential was given. This topic is extremely rich and this overview is therefore by necessity incomplete. Yet, the hope is that with the foundation provided here current research papers will be accessible to a newcomer in the field. In the second part new methods to evade the sign problem altogether were discussed, in particular complex Langevin dynamics. These approaches are still very much in development and hence a conclusion or consensus has not yet been reached, let alone a complete determination of the QCD phase diagram. The upshot of this is that there is still a lot of scope for input, progress and clever ideas, which, combined with the increase of computing power, may eventually  result in {\em the} QCD phase diagram.

\section*{Acknowledgements}

It is a pleasure to thank the organisers -- especially Marcelo Chiapparini --, the participants and the other lecturers for the excellent meeting in a pleasant environment. Special thanks go to Awena Jones.
Over the past years, I have learnt about QCD at nonzero chemical potential from many people. In particular I would like to thank Simon Hands, Nucu Stamatescu,  Erhard Seiler, D\'enes Sexty, Kim Splittorff, Philippe de Forcrand and Owe Philipsen. 
My work is supported by STFC under grant ST/L000369/1, the Royal Society and the Wolfson Foundation.

\appendix

\section{Relativistic Bose gas at nonzero chemical potential}
\label{App:A}

Consider a self-interacting complex scalar field in the presence of a 
chemical potential $\mu$, with the continuum action
 \be
 S = \int d^4x\,\left[ |\partial_\nu\phi|^2
 + (m^2-\mu^2)|\phi|^2 
 + \mu\left(\phi^*\partial_4\phi - \partial_4\phi^* \phi \right)
 + \lambda|\phi|^4 \right].
\label{eq11}
\ee
 The euclidean action is complex and satisfies $S^*(\mu) = S(-\mu^*)$. Take $m^2>0$, so that at 
 vanishing  and small $\mu$ the theory is in its symmetric phase.

The lattice action, with lattice spacing $a_{\rm  lat}\equiv 1$,  is
\be
 S = \sum_x \bigg[ \left(2d+m^2\right) \phi_x^*\phi_x 
 + \lambda\left( \phi_x^*\phi_x\right)^2
- \sum_{\nu=1}^4\left(  \phi_x^* e^{-\mu\delta_{\nu,4}} \phi_{x+\hat\nu} 
+ \phi_{x+\hat\nu}^* e^{\mu\delta_{\nu,4}} \phi_x \right)
\bigg],
\ee
where the number of euclidean dimensions is $d=4$. 

\noindent {\em i}) Show that this action reduces to (\ref{eq11}) in the continuum limit.

\noindent {\em ii})
The complex field is written in terms of two real fields $\phi_a$ ($a=1,2$) as $\phi=\frac{1}{\sqrt{2}}(\phi_1+i\phi_2)$. Show that the lattice action then reads
 \bea
 S 
 =&&\hm \sum_x\bigg[ \half\left(
 2d+m^2\right) \phi_{a,x}^2
 + \frac{\lambda}{4}\left(\phi_{a,x}^2\right)^2
 - \sum_{i=1}^3 \phi_{a, x}\phi_{a, x+\hat i}
 \nn \\ && \hm\hspace*{1.5cm}
 -\cosh\mu\,  \phi_{a, x}\phi_{a, x+\hat 4}
 +i\sinh\mu\, \vareps_{ab}\phi_{a, x}\phi_{b, x+\hat 4}
\bigg],
\label{eq:S}
\eea
 where  $\vareps_{ab}$ is the antisymmetric tensor with $\eps_{12}=1$, and summation over repeated indices is implied.
Note that the `$\sinh\mu$' term is complex.

From now on the self-interaction is ignored and we take $\lambda=0$.
 After going to momentum space,  the action (\ref{eq:S}) reads
 \be
 \label{eq:S0}
S = \sum_p  \half\phi_{a,-p}\left(\delta_{ab}A_p 
-\vareps_{ab}B_p\right)\phi_{b,p}
= \sum_p\half \phi_{a,-p} M_{ab,p} \phi_{b,p},
\ee
where
\be
 M_p = \left(
\begin{array}{cc}
A_p & -B_p \\
B_p & A_p 
\end{array}
\right),
\ee
and
 \be
A_p= m^2 + 4\sum_{i=1}^3\sin^2\frac{p_i}{2}  + 2 \left( 1- 
\cosh \mu \cos p_4\right),
\;\;\;\;\;\;
B_p = 2 \sinh\mu \sin p_4.
\label{eq:AB}
\ee

\noindent {\em iii})
 Show that the propagator  corresponding to the action (\ref{eq:S0}) is
 \be
 G_{ab,p} = \frac{\delta_{ab} A_p + \vareps_{ab} B_p}{A_p^2+B_p^2}.
 \ee
 
 \noindent {\em iv})
 Demonstrate that the dispersion relation that follows from the poles of the propagator, 
taking $p_4=iE_\pv$, reads
\be
 \cosh E_\pv(\mu) = \cosh\mu\left(1+\half \hat \om_\vecp^2\right) \pm \sinh\mu 
\sqrt{ 1+\frac{1}{4} \hat \om_\vecp^2},
\ee
where 
\be
 \hat \om_\vecp^2 = m^2+  4\sum_i\sin^2\frac{p_i}{2}.
\ee

\noindent {\em v})
Show that this can be written as
\be
 \cosh E_\pv(\mu) = \cosh \left[E_\pv(0) \pm \mu\right],
\ee
such that the (positive energy) solutions are 
\be
 E_\pv(\mu) = E_\pv(0) \pm \mu.
\ee
Sketch the spectrum. Note that the critical $\mu$ value for onset  is $\mu_c=E_\vecnul(0)$, so 
that one mode becomes exactly massless at the transition (Goldstone boson).

\noindent {\em vi})
The phase-quenched theory corresponds to $\sinh\mu=B_p=0$. Show that the dispersion relation in the phase-quenched theory is 
\be
 \cosh E_\pv(\mu) = \frac{1}{\cosh\mu}\left(1+\half \hat 
\om_\vecp^2\right),
\ee
 which corresponds to $E_\pv^2(\mu) = m^2-\mu^2+\pv^2$ in the continuum limit. 

\noindent {\em vii})
Compare the spectrum of the full and the phase-quenched theory, when $\mu<\mu_c$.
 At larger $\mu$, it is necessary to include the self-interaction to stabilize the theory.  Based on what you know about symmetry breaking, sketch the spectrum in the full and the phase-quenched theory at larger $\mu$  as well.

Although the spectrum depends on $\mu$, thermodynamic quantities do not.
Up to an irrelevant constant, the logarithm of the partition function is
\be
\ln Z = -\half\sum_p\ln\det M = -\half\sum_p\ln(A_p^2+B_p^2),
\ee  
and some observables are given by
\be
\label{eq:phi2}
\bra|\phi|^2\ket = -\frac{1}{\Omega}\frac{\partial \ln Z}{\partial 
m^2} = \frac{1}{\Omega}\sum_p\frac{A_p}{A_p^2+B_p^2},
\ee
and
\be
\label{eq:dens2}
\bra n\ket = \frac{1}{\Omega}\frac{\partial \ln Z}{\partial \mu}
= 
 -\frac{1}{\Omega}\sum_p \frac{A_pA_p'+B_pB_p'}{A_p^2+B_p^2},
\ee
where  $\Omega=N_\sigma^3N_\tau$ and $A'_p=\partial A_p/\partial\mu$,  $B'_p=\partial 
B_p/\partial\mu$.

 \noindent {\em viii})
Evaluate the sums (e.g.\ numerically) to demonstrate that thermodynamic quantities are independent of $\mu$ in the thermodynamic limit at vanishing temperature.
 
 \noindent
 This exercise is based on Ref.\  \cite{Aarts:2009hn}.

\section{One-dimensional QCD}
\label{App:B}

Consider QCD in one (temporal) dimension, with the staggered fermion action
\be
S = \sum \bar\chi (D+m)\chi 
=  \sum_{x=1}^n \left[ \half\bar\chi_x e^{\mu}U_{x,x+1}\chi_{x+1} - \half\bar\chi_{x+1}e^{-\mu}U^\dagger_{x,x+1}\chi_x+ m\bar\chi_x\chi_x\right].
\ee
Here $n$ denotes the number of points in the time direction and is taken to be even. The quarks obey anti-periodic boundary conditions. The links $U_{x,x+1}$ are elements of U($N$) or SU($N$) and transform as $U_{x,x+1}\to \Omega_x U_{x,x+1}\Omega_{x+1}^\dagger$.

Via a unitary transformation, all links but one can be transformed away (``temporal gauge''), i.e.\ $U_{n,1}\equiv U$, all other $U$'s are unity. The determinant can then be written, up to an overall constant, as \cite{Bilic:1988rw,Ravagli:2007rw}
\be
\det(D+m) = \det\!{}_C\!\left( e^{n\mu_c}+e^{-n\mu_c}+e^{n\mu}U+e^{-n\mu}U^\dagger\right).
\ee
The remaining determinant is in colour space and $\mu_c$ is related to the mass $m$ as
\be
m=\sinh\mu_c.
\ee
The reason for introducing $\mu_c$ will become clear below.

\noindent {\em i}) Show that the determinant has the usual symmetry under complex conjugation.

\noindent {\em ii})
In one dimension, the partition function is simply
\be
Z_{N_f} = \int dU \, \det\!{}^{N_f}\!\left(D+m\right),
\ee
since there is no Yang-Mills action. From now on we take as gauge group U(1): this captures all the essential characteristics in one dimension but also allows one to do the group integral without any effort. We hence write
\be
U=e^{i\phi}, \qquad\quad\quad \int dU = \int_0^{2\pi}\frac{d\phi}{2\pi}.
\ee

\noindent  Show that the partition function for $N_f=2$ is independent of $\mu$ and equal to
\be
Z_{N_f=2} = 4+2\cosh(2n\mu_c).
\ee
Note that the $\mu$ independence is generic in U($N$) theories, since $\mu$ can be absorbed in the U(1) phase (take $\mu$ to be imaginary for this). This is of course not possible in SU($N$) theories, where there is no such freedom.

\noindent {\em iii})  Show that the phase-quenched $N_f=2$ partition function depends on $\mu$ and equals
\bea
\nn
Z_{N_f=1+1^*} =&&\hm  \int dU\, \left|\det(D+m)\right|^2 = \int dU\, \det(D(\mu)+m)\det(D(-\mu)+m) \\
=&&\hm  2+2\cosh(2n\mu_c)+2\cosh(2n\mu).
\eea

\noindent  {\em iv}) 
The chiral condensate and the number density are defined by
\be
\Sigma = \frac{1}{n}\frac{\partial\ln Z}{\partial m},
\qquad\quad\quad
\bra n_B\ket = \frac{1}{n}\frac{\partial\ln Z}{\partial \mu}.
\ee

\noindent Show that in the full theory one finds
\be 
\Sigma = \frac{2\sinh(2n\mu_c)}{2+\cosh(2n\mu_c)}\frac{1}{\cosh\mu_c} \to \frac{2\sgn(\mu_c)}{\cosh\mu_c},
\qquad\quad\quad
\bra n_B\ket = 0.
\ee
The arrow denotes the thermodynamic limit $n\to\infty$. The $\mu$ independence is obvious.

\noindent {\em v}) Show that in the phase-quenched theory one finds on the other hand
\be 
\Sigma = \frac{2\sinh(2n\mu_c)}{1+\cosh(2n\mu_c)+\cosh(2n\mu)}\frac{1}{\cosh\mu_c} 
 \to
 \begin{cases}
 \frac{2\rm{sgn}(\mu_c)}{\cosh\mu_c}&  |\mu|<|\mu_c|\\
0 & |\mu|>|\mu_c|
\end{cases},
\ee
and
\be
\bra n_B\ket = \frac{2\sinh(2n\mu)}{1+\cosh(2n\mu_c)+\cosh(2n\mu)}
 \to 
 \begin{cases}
 0 &  |\mu|<|\mu_c|\\
2\sgn(\mu) & |\mu|>|\mu_c|
\end{cases}.
\ee
The full and phase-quenched theories agree when $\mu<\mu_c$ (no $\mu$ dependence). The phase-quenched theory undergoes a phase transition at $\mu=\mu_c$, where the density jumps to 2. The interesting region in view of the Silver Blaze problem is therefore this large $\mu$ region, where the sign problem is severe and the average phase factor  vanishes in the thermodynamic limit:
\be
\bra e^{2i\varphi}\ket_{\rm pq} = \frac{Z_{N_f=2}}{Z_{N_f=1+1^*}} \to 0,
\qquad\quad\quad \det(D+m) = e^{i\varphi}|\det(D+m)|.
\ee

\noindent {\em vi})
The eigenvalues of $D$ are
\be
\lambda_k = \half e^{i(2\pi(k+\half)+\phi)/n+\mu} - \half e^{-i(2\pi(k+\half)+\phi)/n-\mu}
\quad\quad\quad (k=1,\ldots,n).
\ee
The $k+\half$ arises from the antiperiodic boundary conditions and the $\phi/n$ from uniformly distributing the link $U$ over all links as $U^{1/n}$.

\noindent  Demonstrate that the eigenvalues lie on an ellipse in the complex plane, determined by 
\be
\left(\frac{\re\lambda_{k}}{\sinh(\mu)}\right)^2
+
\left(\frac{\im\lambda_{k}}{\cosh(\mu)}\right)^2 =1.
\ee
The transition in the phase-quenched theory occurs when the quark mass gets inside this ellipse.

\noindent {\em vii}) 
To compute the eigenvalue density,
\be
\rho(z;\mu)= \frac{1}{Z_{N_f}}\int dU \, \det\!{}^{N_f} (D+m) \, \sum_k  \delta^2(z- \lambda_k),
\ee
we therefore parametrize
\be
z = \half\left( e^{i\alpha+\mu} - e^{-i\alpha-\mu}\right),
\ee
such that
\be
\label{eqs}
\Sigma = \int_0^{2\pi}\frac{d\alpha}{2\pi}\,\frac{\rho(\alpha;\mu)}{z(\alpha)+m}.
\ee
One then finds, for $N_f=2$,
\be
\rho(\alpha;\mu) = \frac{4\left[\cosh(n\mu_c)+\cosh(n(\mu+i\alpha))\right]^2}{2+\cosh(2n\mu_c)}.
\ee

\noindent Show that in the thermodynamic limit, the eigenvalue density behaves as
\be
\rho(\alpha;\mu) = 
 \begin{cases}
 2 &  |\mu|<|\mu_c|\\
2e^{2n\left(|\mu|-|\mu_c|+i\alpha\right)} & |\mu|>|\mu_c|
\end{cases},
\ee
i.e.\ it is well-behaved when the full and phase-quenched theories agree, but it is complex and oscillating with a divergent amplitude in the Silver Blaze region.

\noindent {\em viii}) Show that these oscillations are necessary to find a $\mu$ independent chiral condensate by evaluating Eq.\ (\ref{eqs}) explicitly (write $e^{i\alpha}=w$ and use contour integration).

 \noindent
 One-dimensional QCD is discussed in Refs.\ \cite{Gibbs:1986xg,Bilic:1988rw,Ravagli:2007rw}. This exercise is based on Ref.\ \cite{Aarts:2010gr}.

\section{Fokker-Planck equation}
\label{App:C}

Consider the Langevin process
\be
\dot x(t) = K[x(t)]+\eta(t), \quad\quad\quad K(x)=-S'(x),  \quad\quad\quad 
\bra\eta(t)\eta(t')\ket_\eta=2\lambda\delta(t-t'),
\ee
where $\lambda$ normalizes the noise. 
We want to derive the associated Fokker-Planck equation
\be
\label{eqFP}
\partial_t \rho(x,t) = \partial_x\left(\lambda\partial_x-K\right)\rho(x,t),
\ee
for the distribution $\rho(x,t)$, defined via 
\be
\label{eq1}
\bra O[x(t)]\ket_\eta = \int dx\,\rho(x,t)O(x),
\ee
with $O(x)$ a generic observable.
Here the subscript $\eta$ denotes noise averaging and  will be dropped from now on.

To achieve this we consider the discretized process
\be
\delta_n\equiv x_{n+1} - x_n =\eps K_n+\sqrt{\eps}\eta_n, \qquad\quad\quad  \bra\eta_n\eta_{n'}\ket=2\lambda\delta_{nn'}.
\ee

\noindent {\em i}) Show that
\bea
\nn
\bra O(x_{n+1})\ket - \bra O(x_n)\ket =&&\hm  \bra O'(x_n)\delta_n + \half O^{''}(x_n)\delta_n^2 + \ldots\ket
\\
=&&\hm \eps \bra O'(x_n) K_n + \lambda O^{''}(x_n)\ket +{\cal O}(\eps^{3/2}).
\eea

\noindent In the $\eps\to 0$ limit, this gives
\be
\partial_t \bra O(x)\ket =\bra O'(x) K(x) + \lambda O^{''}(x)\ket.
\ee

\noindent {\em ii}) Use Eq.\ (\ref{eq1}) to demonstrate that this yields the Fokker-Planck equation (\ref{eqFP}) for $\rho(x,t)$. What should  $\lambda$ be in order to obtain the desired equilibrium distribution?

\noindent {\em iii)} We now repeat the analysis for the complex Langevin equations, 
\bea
\nn
&&\hm  \dot x= K_x+\eta_x, \qquad\quad  K_x=-\re\, S'(z),  \qquad\quad 
\bra\eta_x(t)\eta_x(t')\ket=2\lambda_x\delta(t-t'), \\
&&\hm \dot y = K_y+\eta_y, \qquad\quad  K_y=-\im\, S'(z),  \qquad\quad 
\bra\eta_y(t)\eta_y(t')\ket=2\lambda_y\delta(t-t').
\eea
By writing $z=x+iy$, show that these Langevin equations are equivalent to 
\be
 \dot z= -S'(z)+\eta, \qquad\quad\quad \bra\eta(t)\eta(t')\ket=2\delta(t-t').
\ee
Express $\eta$ in terms of $\eta_{x,y}$ and derive the necessary restrictions on $\lambda_{x,y}$ (answer: $\lambda_x-\lambda_y=1$).
The case $\lambda_y>0$ is referred to as complex noise.

\noindent {\em iv)} The distribution $P(x,y;t)$ is now defined via
\be
\bra O[x(t)+iy(t)]\ket_\eta = \int dxdy\, P(x,y;t)O(x+iy).
\ee
Show that $P(x,y;t)$ satisfies
\be
\partial_t P(x,y;t) = \left[\partial_x\left(\lambda_x\partial_x - K_x\right) + \partial_y\left(\lambda_y\partial_y - K_y\right) \right] P(x,y;t).
\ee
The case $\lambda_x=1$, $\lambda_y=0$ is used in the main text.

\noindent
This is reviewed e.g.\ in Ref.\ \cite{Damgaard:1987rr}. Complex noise and especially its problems are discussed in Ref.\ \cite{Aarts:2009uq}.

\section{Yet another Gaussian model}
\label{App:D}

\noindent
Consider the complex integral
\be
Z = \int_{-\infty}^\infty dx\,\rho(x), \quad\quad\quad \rho(x) = e^{-S}, \quad\quad\quad S=\half\sigma x^2, \quad\quad\quad \sigma=a+ib.
\ee

\noindent {\em i}) Show that the corresponding complex Langevin equations are given by
\begin{align}
&\dot x = K_x +\eta, & K_x=-ax+by, \\
&\dot y = K_y, & K_y=-ay-bx,
\end{align}
where $\bra\eta(t)\eta(t')\ket=2\delta(t-t')$.

\noindent {\em ii}) Demonstrate that these Langevin equations are solved by
\bea
 x(t) =&&\hm e^{-at}
\left[\cos(bt)x(0) +\sin(bt)y(0)\right] + \int_0^t ds\,e^{-a(t-s)}\cos[b(t-s))]\eta(s),
\\
 y(t) =&&\hm e^{-at} \left[\cos(bt)y(0) -\sin(bt)x(0)\right]
- \int_0^t ds\,e^{-a(t-s)} \sin[b(t-s)]\eta(s).
\eea

\noindent {\em iii}) Show that the expectation values in the infinite time limit are given by
\be
 \bra x^2\ket= \frac{1}{2a}\frac{2a^2+b^2}{a^2+b^2}, \quad\quad
 \bra y^2\ket= \frac{1}{2a}\frac{b^2}{a^2+b^2}, \quad\quad
 \bra xy\ket= -\frac{1}{2}\frac{b}{a^2+b^2}.
\ee

\noindent {\em iv}) Demonstrate that this yields the desired result
\be
\bra x^2\ket \to \bra (x+iy)^2\ket = \frac{a-ib}{a^2+b^2} 
= \frac{1}{a+ib} =\frac{1}{\sigma}.
\ee

\noindent  {\em v})
The Fokker-Planck equation for the (real and positive) weight $P(x,y;t)$, defined via
\be
\bra O(x(t)+iy(t))\ket = \int dxdy\, P(x,y;t) O(x+iy),
\ee
 is given by
\be
\partial_t P(x,y;t) = 
\left[  \partial_x \left( \partial_x-K_x\right) - \partial_y K_y \right] P(x,y;t)
\ee
Since the original integral is Gaussian, the equilibrium distribution $P(x,y)$ is also Gaussian and can be written as
\be
P(x,y) = N\exp\left[ -\alpha x^2-\beta y^2-2\gamma xy\right],
\ee
where $N$ is a normalization constant.

\noindent Using the Fokker-Planck equation, show that the coefficients are given by
\be
 \alpha = a, \quad\quad
 \beta = a\left(1+\frac{2a^2}{b^2}\right), \quad\quad
 \gamma = \frac{a^2}{b},
\ee
and demonstrate that this gives the previously computed expectation values
\be
\bra x^2\ket = \frac{\int dxdy\, P(x,y) x^2}{\int dxdy\, P(x,y)},
\ee
etc.

\noindent {\em vi}) From the equivalence
\be
\int dx\, \rho(x) O(x) = \int dxdy\, P(x,y)O(x+iy),
\ee
it follows that the real distribution is related to the original complex one via
\be
\rho(x) = \int dy\, P(x-iy,y).
\ee
Verify this explicitly (up to the undetermined normalization).

\noindent
This is simple version of the problem treated in Ref.\  \cite{Aarts:2009hn}.

\end{document}